\RequirePackage{fix-cm}
\documentclass[smallextended]{svjour3}       
\smartqed  
\usepackage{graphicx}
\usepackage{cite}
\usepackage{multicol}
\usepackage{subfigure}
\usepackage{graphicx}
\usepackage{float}
\usepackage{caption}
\usepackage{setspace}
\usepackage{amsmath,amssymb,amsfonts}
\usepackage{algorithmic}
\usepackage{textcomp}
\usepackage{makecell}
\usepackage[utf8]{inputenc}
\usepackage{tabularx}
\usepackage{multirow}
\DeclareMathOperator*{\argminA}{arg\,min} 
\DeclareMathOperator*{\argmaxA}{arg\,max} 
\newcommand* \Prr {\mathop{}\!\mathrm{Pr}}

\begin{document}

\title{A Survey: Non-Orthogonal Multiple Access with Compressed Sensing Multiuser Detection for mMTC
}


\author{Mehmood Alam   \and
        Qi Zhang 
}


\institute{M.Alam \at
             Aarhus University, Denmark\\
             \email{Mehmood.Alam@eng.au.dk}           
           \and
           Q. Zhang \at
             Aarhus University, Denmark\\
     \email{qz@eng.au.dk}           
}

\date{Received: date / Accepted: date}

\maketitle

\begin{abstract}
One objective of the 5G communication system and beyond is to support massive machine type of communication (mMTC) to propel the fast growth of diverse Internet of Things use cases. The mMTC aims to provide connectivity to tens of billions sensor nodes. The dramatic increase of sensor devices and massive connectivity impose critical challenges for the network to handle the enormous control signaling overhead with limited radio resource. Non-Orthogonal Multiple Access (NOMA) is a new paradigm shift in the design of multiple user detection and multiple access. NOMA with compressive sensing based multiuser detection is one of the promising candidates to address the challenges of mMTC. The survey article aims at providing an overview of the current state-of-art research work in various compressive sensing based techniques that enable  NOMA. We present characteristics of different algorithms and compare their pros and cons, thereby provide useful insights for researchers to make further contributions in NOMA  using compressive sensing techniques.
\keywords{Non-orthogonal CDMA\and massive connectivity\and grant free medium access\and compressive sensing multiuser detection}
\end{abstract}

\section{Introduction}
\label{sec:introduction}
The fifth generation (5G) wireless communication system envisions three major use cases: (i) enhanced mobile broadband (eMBB),  (ii) ultra-reliable and  low latency communication (URLLC), and (iii)  massive machine type communication (mMTC). The enhanced mobile broadband is characterized by ubiquitous coverage with peak data rate of 20 Gbps and a latency of less than 10 ms \cite{7503829}. The ultra-reliable and low latency communication is focused on enabling a variety of mission critical applications and tactile Internet applications, such as autonomous vehicle, remote industrial control and remote manufacturing with a latency requirement of less than 1 millisecond \cite{zhang2015mission}.  The mMTC aims at providing connectivity to massive  low power and low data rate Internet of Things (IoT)  devices  which will open up  enormous business opportunities in, e.g.,  building automation, smart agriculture, smart cities, fleet management, etc. It is expected that the number of such devices will rise to 50 billion by 2020 \cite{4471881} which will result in the number of connections increased to one million per square kilometer \cite{imt}. In order to accommodate such huge number of devices, the network capacity needs to be significantly improved.\par
\begin{figure}[htbp]
	\centering
	\subfigure[TDMA]{\includegraphics[scale=0.25]{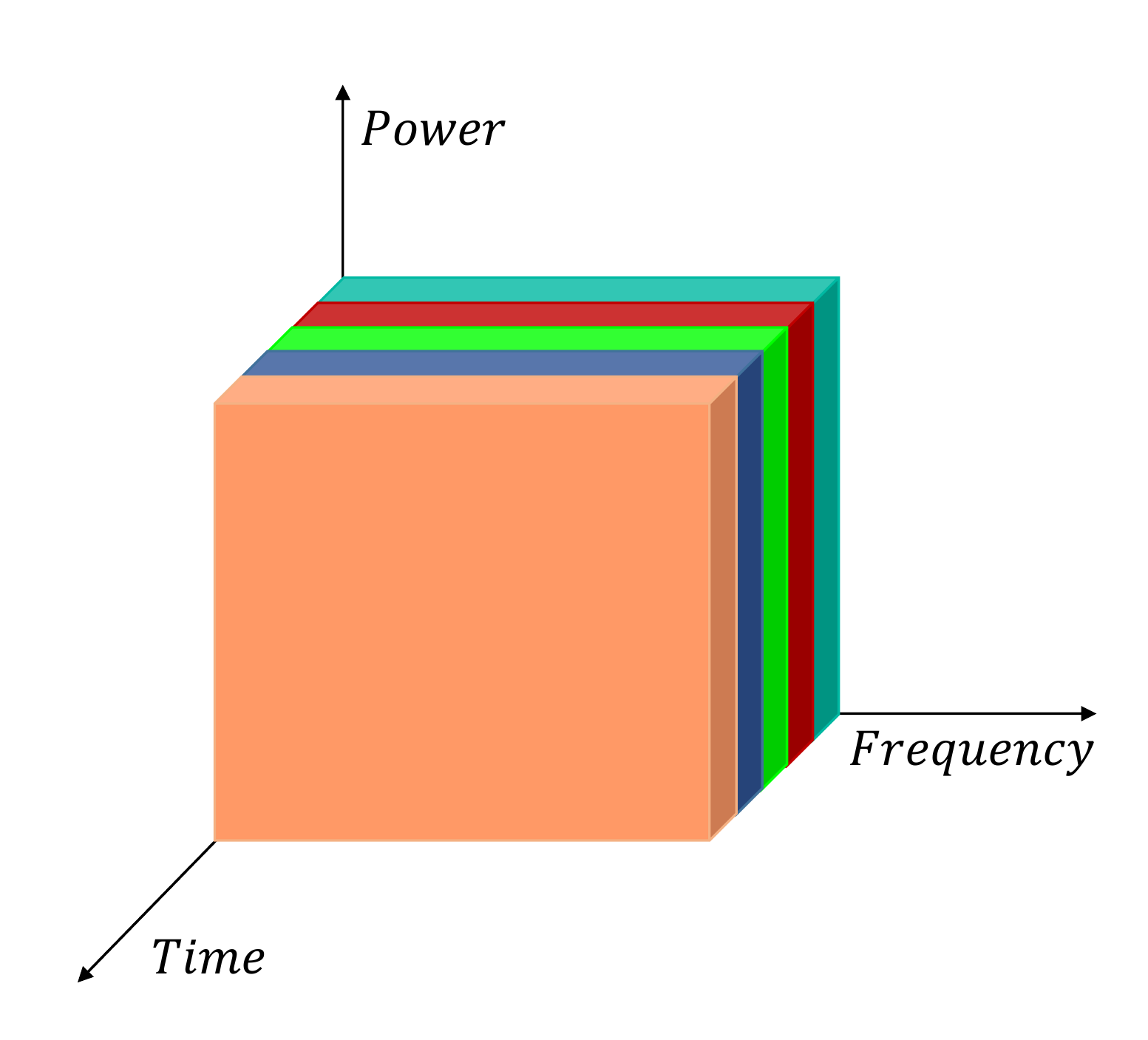}}
	\subfigure[FDMA]{\includegraphics[scale=0.25]{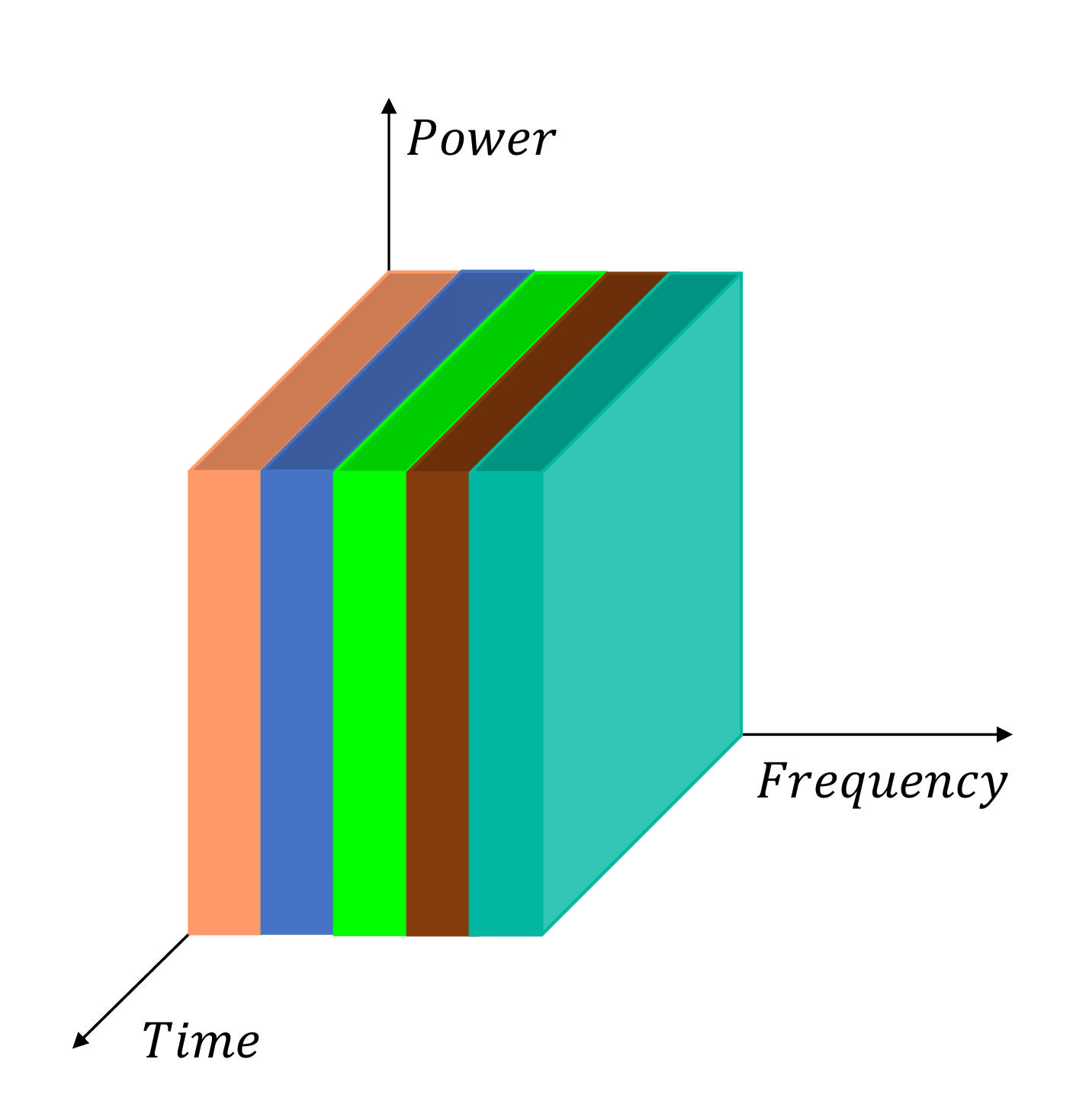}}
	\subfigure[CDMA]{\includegraphics[scale=0.25]{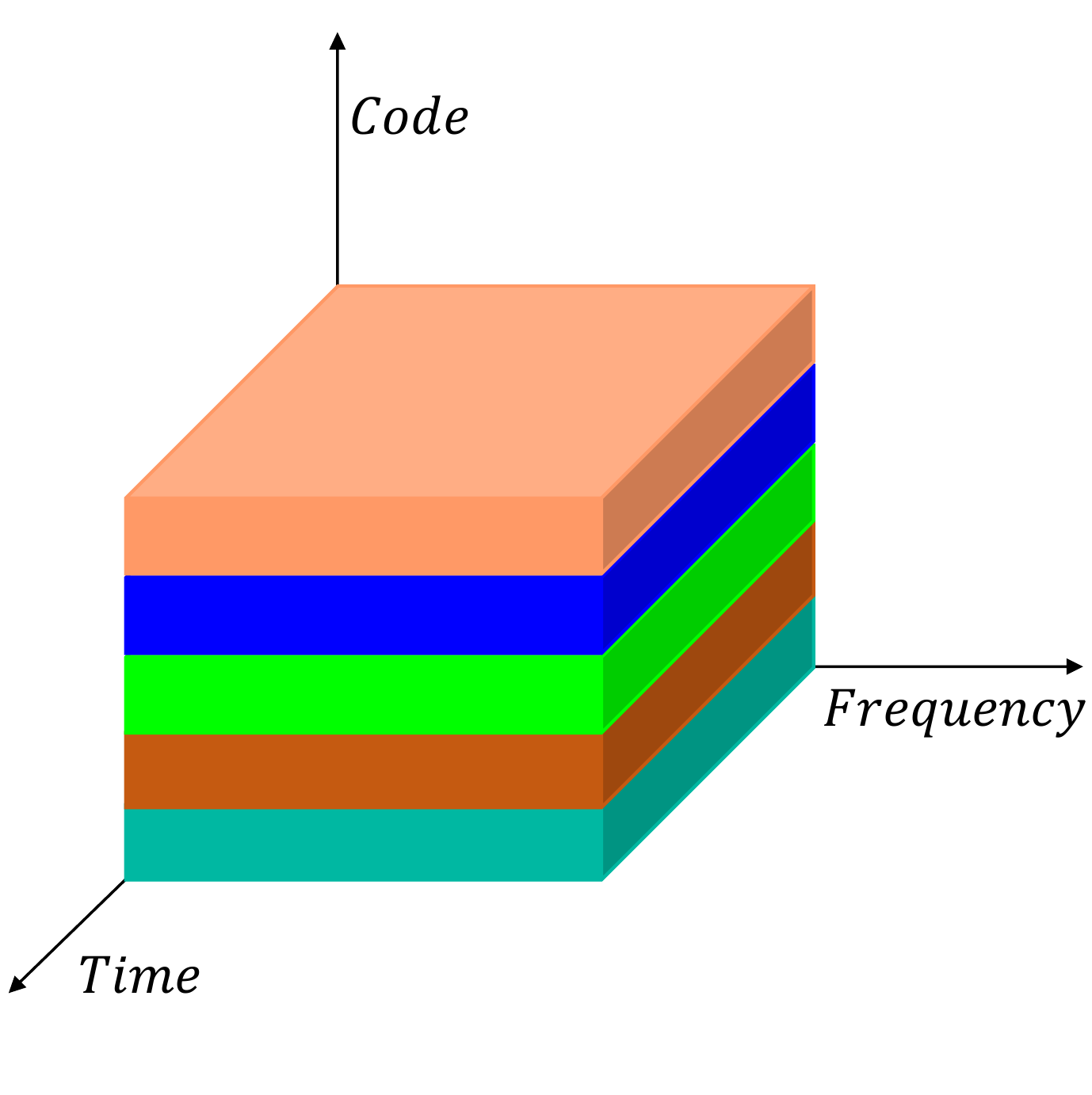}}
	\caption{Resource allocation in orthogonal multiple access schemes: Colors represent different users which are equal to the  number of resources}	\label{omaa}
\end{figure}

 
The current orthogonal multiple access  (OMA)  such as time division multiple access (TDMA), frequency division multiple access (FDMA) and code division multiple access (CDMA)  serves a single user in each orthogonal resource block. Therefore, the maximum number of simultaneously supported devices in an OMA scheme is  limited  by the number of orthogonal resources. For example, in orthogonal CDMA where orthogonal codes are assigned to the users, the maximum number of connected devices cannot exceeds the spreading factor. This orthogonality constraint makes the OMA schemes highly spectral inefficient for mMTC in 5G. Figure \ref{omaa} shows resource allocation in OMA schemes where the resources assigned to the users are orthogonal to each other. Due to this orthogonality constraint, the current OMA based long term evolution (LTE)~\cite{lte}  can only support a fraction of the anticipated number of devices for future mMTC with its control channel element \cite{rx_o1}.
Furthermore, the sporadic transmission of small data packets in mMTC requires minimum  control signaling overhead whereas  LTE has a  high cost of signaling overhead and high channel access latency for small data transmission. Therefore, connecting massive number of resource constrained devices to the network requires a paradigm shift in the multiple access technique. \par 
Non-orthogonal multiple access (NOMA) is a promising technique for massive connectivity which has attracted tremendous interest of researchers from both academia and industry. The NOMA schemes assign non-orthogonal resources to the users and therefore enable system overloading, i.e., allowing users to more efficiently share the same resources. The overloading capability of NOMA scheme is characterized by the overloading factor, which is the ratio of the total number of nodes to the total available orthogonal resources.  At the receiver, advanced multiuser detection (MUD) techniques are deployed to separate the users. The NOMA schemes can be categorized into two main groups: the power domain NOMA and the code domain NOMA.  Note that node and user are used interchangeably in this paper. 
\subsection{Power domain NOMA}
\begin{figure}[h]
	\centering
	\includegraphics[scale=0.3]{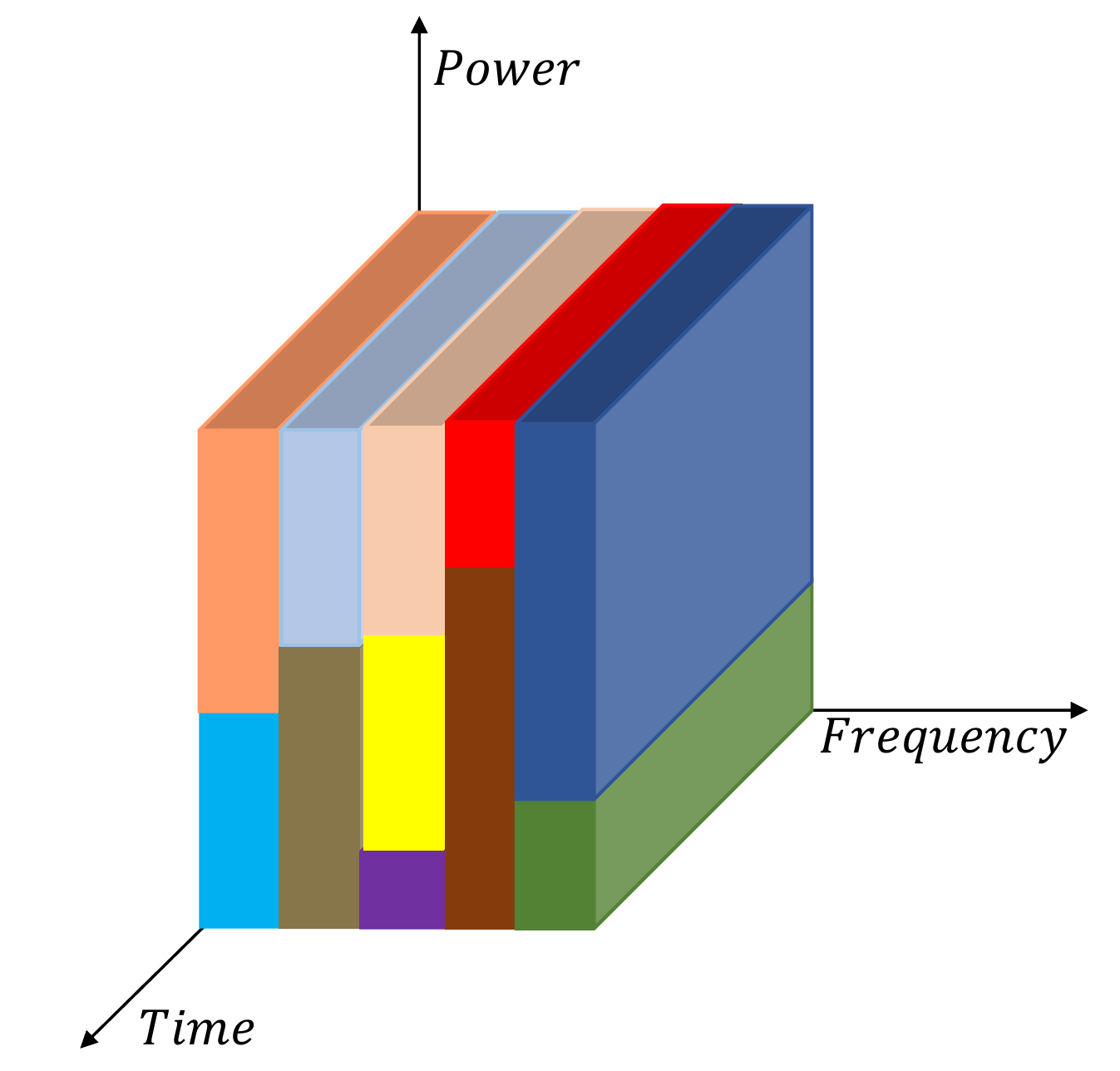}
	\caption{ Resource allocation in power domain NOMA: Colours represent different users which are more than the number of resources} 	\label{nomaa}
\end{figure}

In power domain NOMA, the users are assigned with different power levels which enable them to share the available resources, i.e., time, frequency or code \cite{6704653,7676258}. At the receiver, successive interference cancellation (SIC) is used for multiuser detection, which differentiates the users according to the assigned power levels. The resource allocation in power domain NOMA is shown in Figure \ref{nomaa}. It is depicted that at a given frequency, more than one users can transmit using different power levels.

\begin{figure}[ht]
	\centering
	\includegraphics[scale=.75]{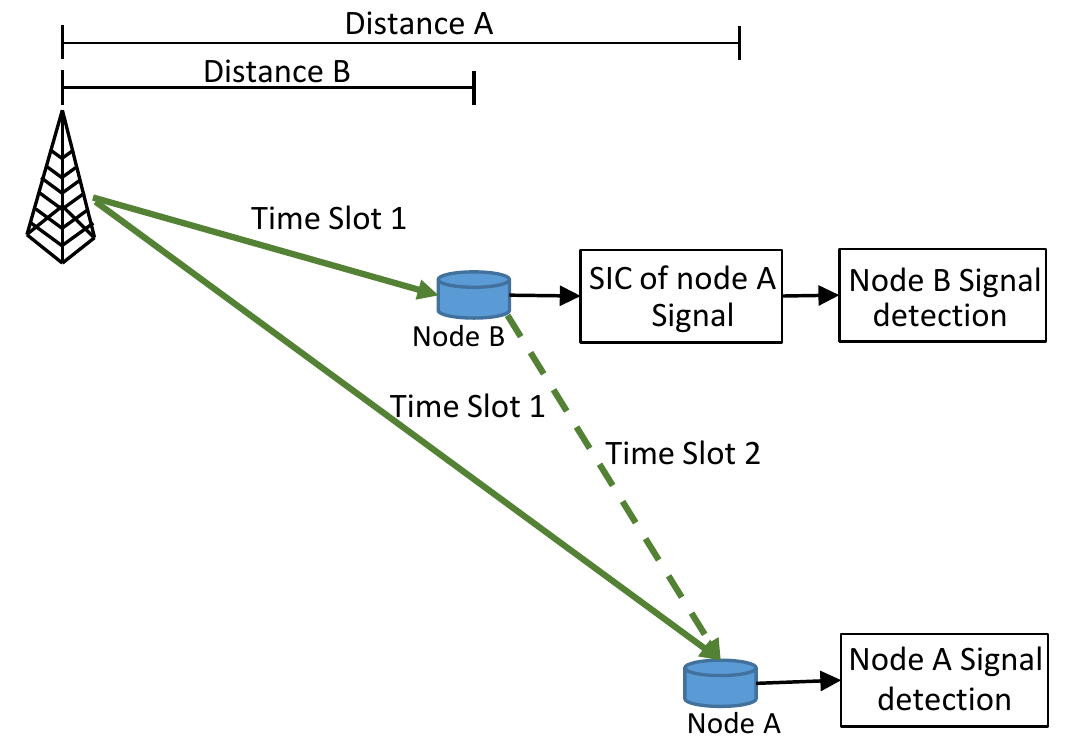}
	\caption{Co-operative power domain NOMA}
	\label{cpnoma}
\end{figure}

It is shown analytically in \cite{6868214} that the NOMA outperforms the OMA techniques in terms of outage performance and ergodic sum rate. However, the user data rate and allocated power should carefully be chosen as the outage performance critically depends on them.  NOMA improves the bandwidth efficiency; however, the fact that the cell-edge users, i.e., the users that are far from the BS, have poor channel conditions than the cell-centered users, i.e., users which are closer to the BS, causes performance degradation to the cell-edge users. This effect is mitigated by using cooperative NOMA \cite{cnoma}. In cooperative NOMA, the users having better channel conditions are paired with users having poor channel conditions.  Figure \ref{cpnoma} depicts a downlink scenario of cooperative NOMA in which a pair of users receives the superimposed signal in the first time slot. In the second time slot the cell-centered user acts as a relay and forwards the received signal to the cell-edge user. The whole task is completed in two time slots using cooperative NOMA which would take three time slots in cooperative OMA \cite{coma}. \par
Cooperative NOMA improves the performance gain of NOMA; however, due to the extra process of relaying, the cell-centered users suffer from battery drainage. In machine type communication, the battery life is of significant importance as the MTC devices are mostly low power and the batteries once installed are rarely changed. To cope with this issue,  in   \cite{nomawet}, the concept of wireless energy transfer, i.e.,  the transfer of energy  from the source to the destination  through the air,  is combined with cooperative NOMA. The proposed protocol is named as cooperative non-orthogonal multiple access with simultaneous wireless information and power transfer (cooperative SWIPT NOMA). In cooperative SWIPT NOMA, the cell-centered users harvest energy from the base station, which is used for the extra step of relaying, hence, reduces the battery drainage. Another variation in power domain NOMA is the introduction of cognitive radio (CR) concept \cite{7273963,7973146}. The conventional power domain NOMA ensures the user fairness; however, it cannot strictly guarantees the user’s quality of service (QoS) targets.  CR-NOMA ensures users QoS targets by designing power allocation policy according to the users QoS requirements. Although the power domain NOMA schemes outperform the OMA schemes, they have higher computational complexity due to the use of SIC. Furthermore, at low SNR the performance gain of power domain NOMA is insignificant \cite{6868214}.\par 
\subsection{Code domain NOMA}
In code domain NOMA, user specific codes are used as multiple access signatures to identify different users at the receiver. The number of available resources are less than the total number of users due to which the codes are non-orthogonal to each other. 
\begin{figure}[h]
	\centering
	\includegraphics[scale=0.3]{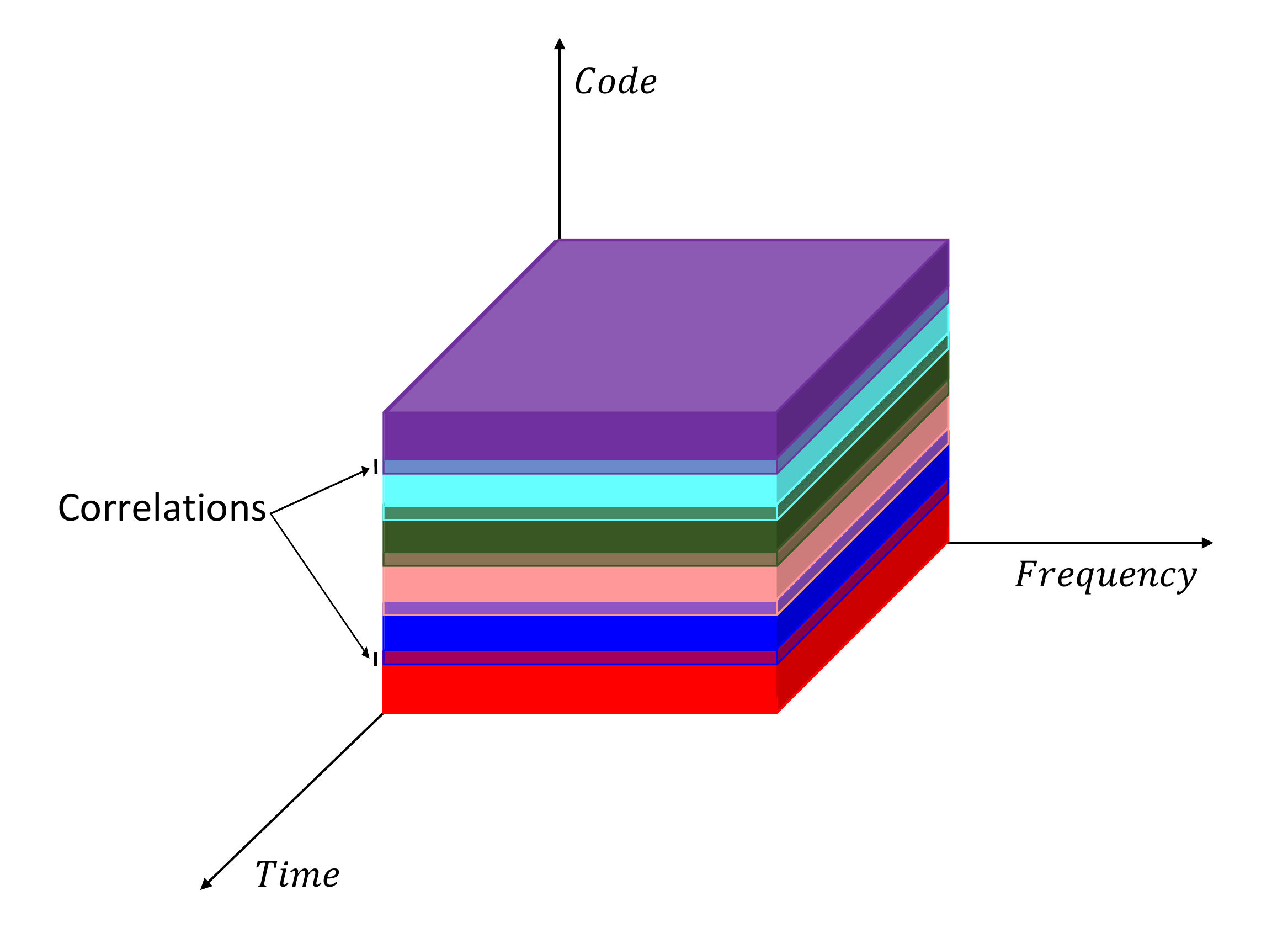}
	\caption{ Resource allocation in code domain NOMA: Colours represent different users which are more than the number of resources} 	\label{nomaac}
\end{figure}
The correlations between the  codes of a non-orthogonal code domain NOMA  is shown in Figure \ref{nomaac} by the overlapping parts of the codes.  The correlations between the codes  increases the probability of errors in  detecting the active users at the receiver. However, advanced multiuser detection techniques such as message passing algorithm, minimum mean square error and SIC are used to efficiently recover the transmitted data. The prominent code domain NOMA schemes include sparse code multiple access (SCMA) \cite{scma,sc1,sc2}, pattern division multiple access (PDMA) \cite{pdma},  multi-user shared access (MUSA) \cite{musa} and so on.  SCMA and PDMA are codebook based schemes, in which a unique codebook is assigned to each user and message passing algorithm (MPA) \cite{mpa} is used at the receiver for multiuser detection. In MUSA the users are separated by assigning low correlated spreading sequence at the transmitter and SIC is used at the receiver for MUD. All these NOMA schemes enable system overloading and hence facilitates the massive connectivity. However, all the nodes within the vicinity of a base station (BS) are assumed active in these schemes while in fact in mMTC a small fraction of the total nodes is active at a time. Moreover, in mMTC the data packet is of small size, for which a grant free medium access control mechanism is indispensable to reduce the control signaling overhead.\par 
To overcome the limitations of the code domain NOMA and to more efficiently exploit the MTC traffic pattern, recently, compressive sensing based multiuser detection (CS-MUD) techniques are used for enabling the code domain NOMA. In order to accommodate large number of nodes within the available resources, non-orthogonal spreading sequences are assigned to the nodes. At the receiver, both the activity and data are jointly detected by using CS-MUD techniques. The CS-MUD exploits the fact that the mMTC transmission is sporadic, i.e., a small number of nodes are active at a time, which allows the compressive sensing techniques to detect the activity at the receiver. The non-orthogonal spreading sequences serves as signatures for the users, which are used to distinguish the active users at the receiver. In  NOMA with CS-MUD, the users transmit their data directly using the spreading sequences, thereby, avoiding the control signaling overhead to access the channel.  \par
\subsection{Contributions}
In the current literature, both the power domain NOMA and the code domain NOMA schemes for 5G are well  investigated and surveyed. The NOMA with CS-MUD is a potential code domain NOMA scheme to meet the mMTC requirements in the 5G wireless system. It has the capability of accommodating massive number of devices with no signaling overhead with comparatively less complex multiuser detection. However, there is no review or survey article, which describes the NOMA with CS-MUD. In this paper, we give an overview of the current research trends within this field and provide useful insights, which will help the researchers to better understand the NOMA with CS-MUD. The   active research areas are classified into three main categories and the progress made in each domain is presented. The first category comprises the development and improvement in designing efficient detection algorithms for the CS-MUD. The second category is the CS based medium access, i.e., how the signatures can efficiently be assigned to the users to gain maximum output in terms of increasing the overloading capability and improving  the activity and data detection. The last category combines CS-MUD with other techniques in order to utilize the advantages of different schemes. Different approaches towards non-orthogonal multiple access with CS-MUD are compared and their pros and cons are explained. Table \ref{cl} categorizes the publications in the literature based on the above-mentioned categories. \par 
\begin{table}[]
	\centering
	\caption{Classification of NOMA with CS-MUD}
	\label{cl}
	\begin{tabular}{|l|l|}
		\hline
		\textbf{Category} 	& \textbf{References} \\ \hline
		CS-Multiuser detection algorithms                    &  \cite{rx_o1,rx_sd1,rx_sd2,rx_kb,rx_np,rx_ga,rx_bols,rx_gait, rx_dn1,rx_dn2,rx_sn1,rx_iorls,rx_mmv,rx_sht,rx_scp}      \\ \hline
		CS-Medium access schemes                       &  \cite{sp_mc1,sp_mc2,sp_mc3,sp_ms1,sp_ms2,sp_mrb,sp_ch1,sp_ch2,sp_oe,o_p1}   \\ \hline
		\hspace{-0.52cm}	\makecell{Combination of CS-MUD with\\ other techniques } &     \cite{o_lds1, o_lds2, o_3,o_mimo1,o_cra1,o_sw}           \\ \hline
	\end{tabular}
\end{table}
The paper is organized as follows. Section \ref{pr} introduces the compressive sensing basics. In Section \ref{noma-cs}, two different ways to formulate the non-orthogonal CDMA with CS-MUD are described.  Section \ref{csalgo} compares the different CS-MUD algorithms and explains their pros and cons. Section \ref{mad} presents the variants of CS based non-orthogonal multiple access schemes and shows their connections and differences. Section \ref{hyb} presents the combination of CS-MUD with other techniques. Section \ref{comparison} compares the representative CS based MUD schemes. Finally Section \ref{con} concludes the paper.\par
\textit{Notations :} In this paper, all boldface uppercase letters represent matrices such as $\mathbf{S}$, while all lowercase boldface letters represent vectors such as $\mathbf{s}$, $\mathbf{x}$. The set of binary, integers and complex numbers are represented by $\mathbb{B}$, $\mathbb{Z}$ and $\mathbb{C}$, respectively.  Italic letters such as $k$, $x$ represent variables. Uppercase letters such as $K$ represents constant values.   
\section{Compressive sensing basics} \label{pr}
Compressed sensing (CS) is a signal processing technique which samples a sparse signal at a rate much less than the Nyquist rate \cite{csss}. A signal $\mathbf{x}\in\mathbb{C}^{K\times1}$ is said to be $K_a$-sparse if it has only $K_a$ non-zero elements, $K_a\ll K$. The signal $\mathbf{{x}}$ can be sparse with respect to any basis $\mathbf{\Phi}$. Let $\mathbf{z}=\mathbf{\Phi x}$ be a compressible signal with respect to $\mathbf{{\Phi}}$. The CS encoding process produces measurement $\textbf{y} \in \mathbb{C}^{N\times 1}$ by a measurement matrix  $\mathbf{\Psi}\in \mathbb{C}^{N\times K}$, $K_a<N<K$,
\begin{equation}
\textbf{y}=\mathbf{\Psi}\textbf{z}+\mathbf{n},
\label{cseq}
\end{equation}
where $\mathbf{n}$ is the background noise vector. The measurement matrix  $\mathbf{{\Psi}}$ and the basis $\mathbf{\Phi}$ should have low coherence  The reconstruction of vector $\mathbf{x}$ is finding a sparse vector $\hat{\mathbf{x}}$ which satisfies Equation  (\ref{cseq}).  As Equation (\ref{cseq}) is an underdetermined system of equations, the reconstruction of vector $\mathbf{x}$ is formulated as
\begin{equation}
\hat{\mathbf{x}}= \argminA_{\mathbf{x} \in \mathbb{C}^{N}} {\Vert \mathbf{x} \Vert}_0   \:\:\:  \text{subject to   }\: \:  \textbf{y} =\mathbf{\Psi} \mathbf{\Phi x},
\label{cs12}
\end{equation}
Where $\Vert.\Vert_0$ is the $l_0$ norm which simply gives the total number of non-zeros elements in the vector. Equation (\ref{cs12}) is a non-convex optimization problem, which is known to be NP-hard to solve. Certain relaxation approaches are used to solve Equation (\ref{cs12}) such as, basis pursuit denoising \cite{chen2001atomic} in which the $l_0$ norm is relaxed to $l_1$ norm. The $l_1$ minimization is a convex optimization problem which recovers the signal from an undermined system of equations with higher accuracy at the cost of higher complexity of cubic order. Another category of compressive sensing reconstruction algorithms is greedy algorithms. In greedy algorithms, e.g., orthogonal matching pursuit (OMP), the support of $\hat{\mathbf{x}}$ is obtained iteratively by selecting the column of $\mathbf{\Psi}$ which has maximum correlation with the residual. The residual is initialized to \textbf{y} and is updated in each iteration of OMP. Once the support is obtained, the corresponding data is estimated by using least square estimation. The greedy algorithms have lower complexity compared to the basis pursuit algorithms at the cost of slightly poor performance.


\section{Non-Orthogonal Multiple Access with CS-MUD} \label{noma-cs}
\begin{figure}[!h]
	\centering
	\includegraphics[scale=.7]{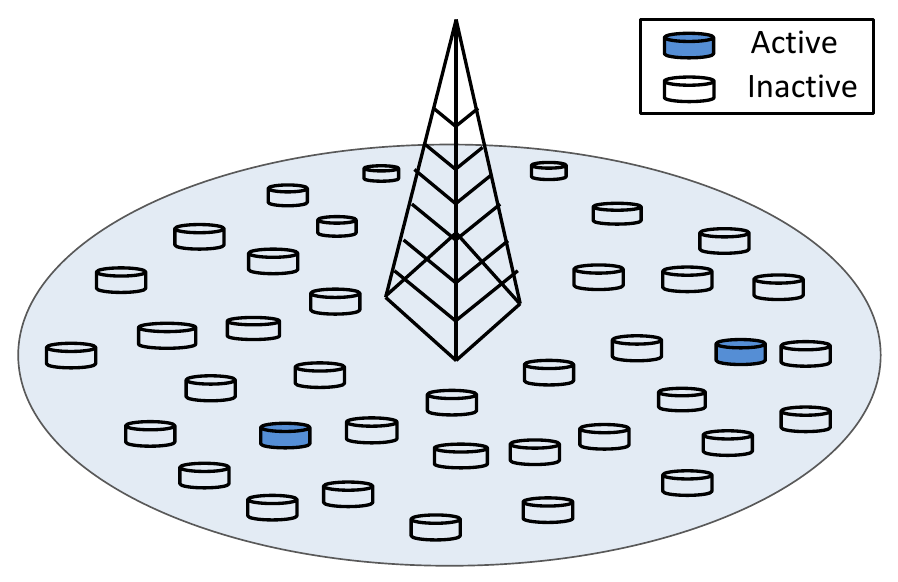}
	\caption{Machine type communication scenario}
	\label{mtc}
\end{figure}

In mMTC the number of simultaneously active users are far less than the total number of users in a cell  \cite{sp_mrb}. A typical uplink mMTC system is depicted in Figure \ref{mtc}, in which a total of $K$ nodes are in the range of a base station (BS) and each node is active with an activity probability, $p_a \ll 1$. Due to this low activity probability, out of $K$ nodes only a small portion of the nodes are simultaneously active. \par 
Compressive sensing multiuser detection exploits this sporadic nature of mMTC to enable non-orthogonal multiple access at the transmitter. In the context of CDMA scheme, the non-orthogonal multiple access is realized by assigning non-orthogonal spreading sequences to the nodes. Two models in the literature are used to formulate the non-orthogonal CDMA as a compressive sensing problem: single measurement vector based compressive sensing (SMV-CS) and multiple measurement vector based compressive sensing (MMV-CS).\par
In SMV-CS model, a one shot transmission is considered. The received signal, $\mathbf{y}$, is a vector which consists of the superimposed symbols of the active nodes.  
\begin{figure}[!h]
	\centering
	\includegraphics[scale=.7]{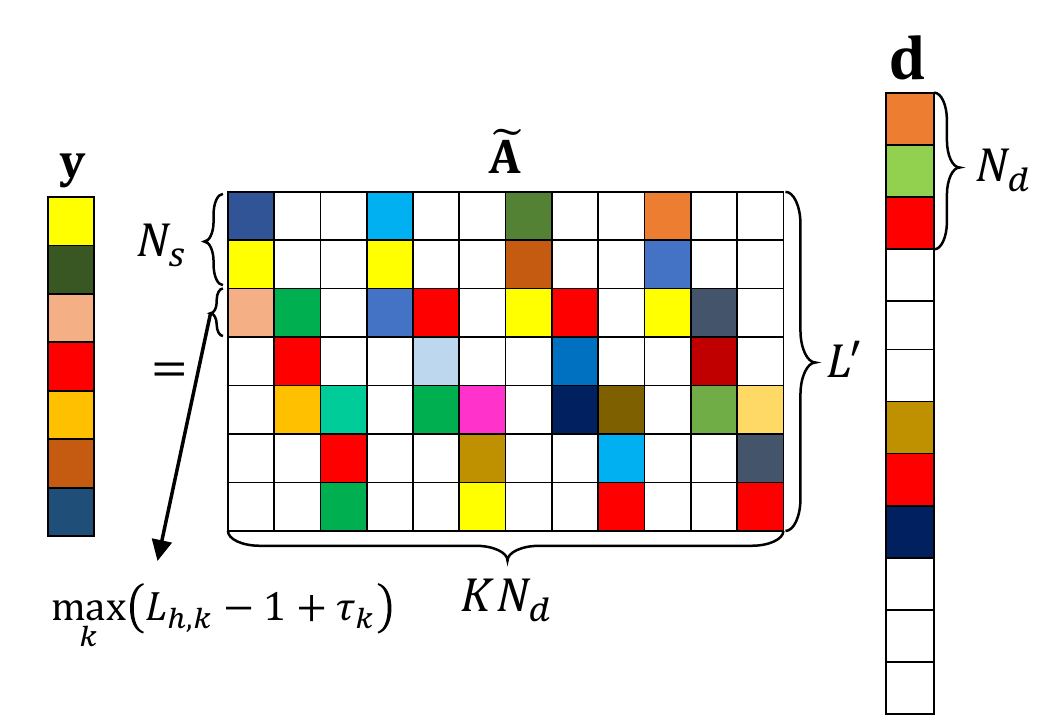}
	\caption{Single measurement vector based compressive sensing model for non-orthogonal multiple access}
	\label{m1}
\end{figure}
Figure \ref{m1} is  a typical SMV-CS representation of an uplink mMTC. The vector $\mathbf{d}$ consists of the data symbols of $K$ users, each with a frame length of $N_d$. Out of $K$ users, only a small fraction, $K_a$, is active. The matrix  $\tilde{\mathbf{A}}\in\mathbb{C}^{{L^\prime} \times KN_d }$ contains the influences of  channel matrix, $\mathbf{H}$, and spreading matrix, $\mathbf{S}$,  where $L^\prime = N_sN_d+\max \limits_{k} ({L_{k,h}}-1+ \tau_k )$, $L_{h,k}$ is the number of channel taps of user $k$,  $\tau_k$ is the relative delay and $N_s$ is the spreading factor \cite{rx_gait}.  In SMV-CS model, when the number of users increases, the size of the sensing matrix, $\tilde{\mathbf{A}}$, becomes huge which leads to poor sampling matrix properties and therefore limits the scalability of the system in terms of hardware cost, memory storage and detection speed. \par
\begin{figure}[!h]
	\centering
	\includegraphics[scale=.7]{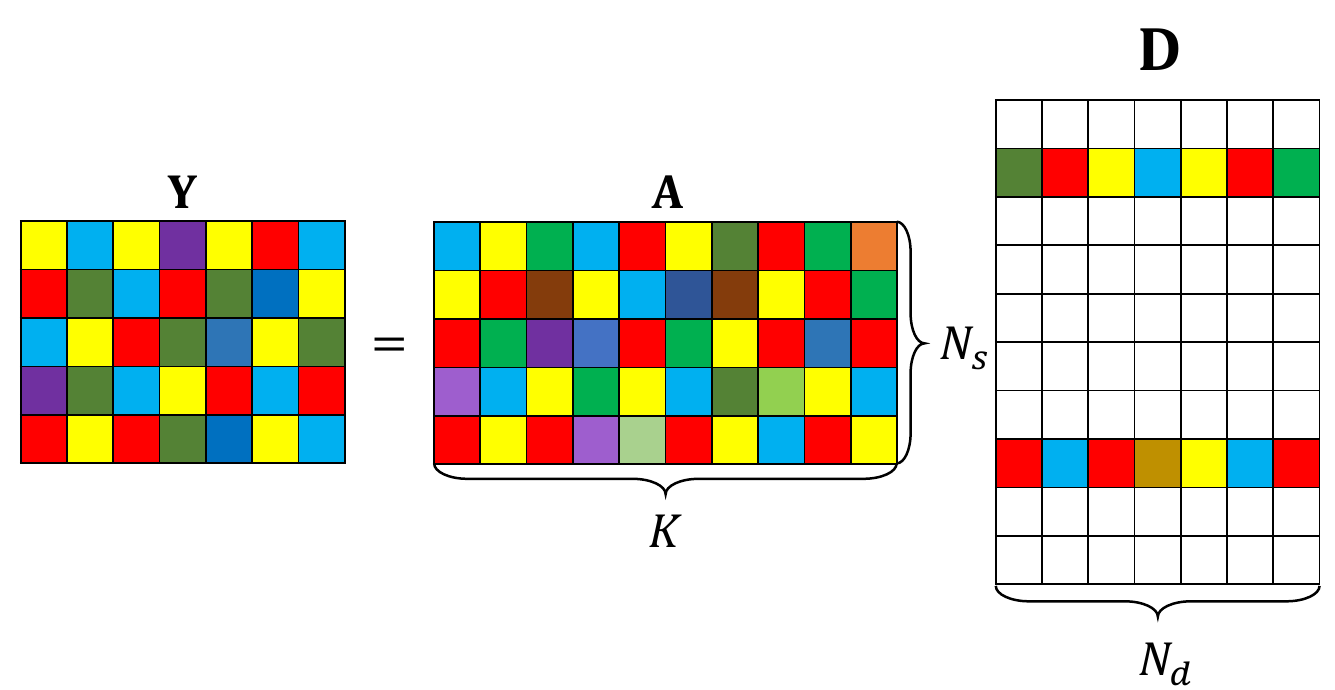}
	\caption{Multiple measurement vector based compressive sensing model for non-orthogonal multiple access}
	\label{m2}
\end{figure}

 
In MMV-CS, the received signal is modeled as a matrix instead of a single vector as shown in Figure \ref{m2}. 
Each row vector of matrix $\mathbf{D}\in\mathbb{C}^{N_s \times N_d}$ in Figure \ref{m2} represents the data frame of a single user and each column vector represents the symbols from all users at time instant, $t$. It is clear that with MMV-CS model, the size of the sensing matrix $\mathbf{A}\in\mathbb{C}^{N_s \times K}$ only depends on the number of users, in other words, independent of the data frame size. Hence, compared with the SMV-CS model, MMV-CS model can better mitigate higher complexity due to the growing number of users. \par 
In \cite{rx_scp,rx_mmv,sp_mc1,sp_mc2,sp_mc3,sp_oe,sp_cb},  MMV-CS is used to represent the non-orthogonal CDMA in uplink mMTC scenario while SMV-CS model is used to describe the system in all other cited papers in this article.
\section{CS-MUD Algorithms} \label{csalgo}
Since the introduction of exploiting the sparsity in mMTC, different CS-MUD algorithms have been proposed. These MUD techniques can be categorized into the following two groups: maximum a posteriori probability (MAP) based algorithms and greedy algorithms. \par 
\subsection{Maximum a posteriori based algorithms}
\subsubsection{Sparsity aware MAP based algorithms} \label{mpb}
In \cite{e1}, for the first time the possibility of exploiting the sporadic nature of user activity for multiuser detection was introduced. For the model 
\begin{equation}
\centering
\mathbf{y}=\tilde{\mathbf{A}}\mathbf{d}+\mathbf{n},
\end{equation}
where $\mathbf{n}\in\mathbb{C}^{N_s\times1}$ is the AWGN noise vector, a sparsity-aware maximum a \textit{posteriori} (S-MAP) criterion is used and the detection process is formulated as \cite{e1}
\begin{equation} \label{smap}
\centering
\hat {\mathbf{d}}^{\text{MAP}}=  \argmaxA_{\mathbf{d}\in\mathcal{A}_a^K} \frac{1}{2}\|\mathbf{y}-\tilde{\mathbf{A}}\mathbf{d}\|_2^2+\lambda\|\mathbf{d}\|_0, 
\end{equation}
where
\begin{equation}
\centering
\lambda := \text{ln} \frac{1-p_a}{p_a/2}
\end{equation}
and $\mathcal{A}$ is the modulation alphabet. Two approaches were followed to solve the detection process in Equation (\ref{smap}): relaxed S-MAP and S-MAP with lattice search. In relaxed S-MAP, it is assumed that  $\mathcal{A}=\{\pm 1,0\} $, which makes Equation (\ref{smap}) equivalent to \cite{e1}
\begin{equation} \label{smap1}
\centering
\hat {\mathbf{d}}^{\text{MAP}}=  \argminA_{\mathbf{{d}}\in\mathcal{A}_a^K} \frac{1}{2}\|\mathbf{y}-\tilde{\mathbf{A}}\mathbf{d}\|_2^2+\lambda\|\mathbf{d}\|_p^p\:,\:\:\: \forall \: p \geq 1. 
\end{equation}\par
Algorithms were designed for  $p=1$ and  $p=2$. Ignoring the finite-alphabet constraint, the optimal solution for $p=1$ is a quadratic programming problem while for $p=2$ it takes a linear form. The relaxed S-MAP multiuser detection is sub-optimal but has advantage of low complexity.  In the second approach, i.e, S-MAP detectors with lattice search, the alphabet $\mathcal A$ is defined as $\mathcal A= \{\pm 1,\pm3,\hdots,\pm(M-1) \}$, with M even. Using the QR decomposition the S-MAP problem is reformulated and the elements of vector $\textbf{d}$  are obtained by searching over a subset of the alphabet. The performance of the detector improves at the cost of higher complexity. The algorithms presented aim at fully loaded ($N_s=K$) CDMA system, however, the lattice search based algorithms can have fair performance for moderate overloaded ($N<K$) case.\par 
In \cite{rx_mmv},  MMV-CS model was considered to reduce the complexity of the receiver and increase the computation speed. The activity is detected based on the covariance matrix of the received signal $\mathbf{Y}\in\mathbb{C}^{N_s\times N_d}$ which is given as \cite{rx_mmv}
\begin{equation}
\Phi_{YY}=\frac{1}{N_d}\mathbf{Y}\mathbf{Y}^H = \mathbf{SV}\mathbf{S}^H + \Phi_{WW},
\label{mmpe}
\end{equation}
where $\Phi_{WW}$  is the sample noise covariance matrix, $\mathbf{S}$ is the spreading matrix and $\mathbf{V}=\text{E}(\mathbf{D} \mathbf{D}^H)$ is the covariance matrix of the transmitted multiuser frame.  The $k$-th user is active if the $k$-th diagonal element of $\mathbf{V}$ is 1.
A MAP detection is used to obtain the positions of non-zero elements in $\mathbf{V}$. The complexity of the proposed algorithms is invariant to the length of the frame, ${N_d}$, however, the performance of the algorithms is dependent on  ${N_d}$ and  improves with increasing ${N_d}$. It is derived that for reliable activity detection the length of the sequence should be greater than the square root of the number of nodes \cite{rx_mmv}. Furthermore, the simulation results show that the algorithms can handle asynchronous frames where the time shifts are up to 60\% of the frame duration.

\subsubsection{Sphere decoding based MUD}
In \cite{rx_sd1}, the   MAP detection for non-orthogonal CDMA is formulated as 
\begin{equation}
\mathbf{\hat{d}} =\argminA_{\mathbf{{d}} \in \mathcal{A}} \Vert\mathbf{y}-\mathbf{Ad}\Vert_2^2-\sigma^2\log \Prr{\{ \mathbf{d} \}},
\label{ssd}
\end{equation}
where $  \Prr \{ \mathbf{d} \}$ is the probability distribution  for the vector $\mathbf{d}$ and $\sigma^2$ is the noise variance.  In \cite{e1} a simple Bernoulli traffic model was considered for the probability distribution of $\mathbf{d}$ in which each user is independently active with a probability $P_a\ll 1$.  To consider a more realistic mMTC traffic, Poisson model was considered instead of Bernoulli in  \cite{rx_sd1} and it is shown that $ \sigma^2 \log \Prr \{ \mathbf{d} \} $ is monotonically increasing. A sorting algorithm is also proposed which reduces the searching levels of the sphere decoding by sorting the sensing matrix according to the correlations with the received signal.  The proposed multiuser detectors give optimal performance; however, the detection complexity is still high when the number of nodes is enormous. Moreover, the detection algorithm exploits the sparsity for joint activity and data detection in a fully loaded system where $N_s=K$ and does not consider the overloaded system.\par 
In CS-MUD, the activity is detected at the receiver and the detection errors, i.e., false alarms and miss detections, lead to significant performance loss. The activity detection therefore becomes a crucial step. In the context of sphere decoding, the mitigation of detection errors is addressed in \cite{rx_np}. The author proposed a Neyman-Pearson \cite{neyman1992problem} based approach to reduce the detection error. Analytically, the problem is formulated as finding an optimal threshold $t^*$ which minimizes the probability of false alarm, $ P_{F_a} $, while keeping the probability of miss detection, $ P_{M_d} $, below a pre-defined threshold, $\eta$ \cite{neyman1992problem}    
\begin{equation}
\centering
t^* = \argminA P_{F_a}(t)  \hspace{0.6 cm} \text{s.t.}\hspace{0.2cm}   P_{M_d}(t) \le \eta.
\label{np}
\end{equation}
The calculation of probabilities in Equation (\ref{np}) is based on the activity log likelihood ratios of the received symbols, which are estimated by using sphere decoding. \par 
\subsubsection{K-best detector for sphere decoding}
The sphere decoder based MUDs are considered to achieve the maximum a posteriori probability (MAP) performance in terms of BER \cite{e1}. However, there are two main disadvantages of sphere decoding, i.e.,  it has an exponential complexity \cite{sd6} and  it is not possible to parallelize the computations \cite{kb16}. In \cite{rx_kb} to address these issues, a K-best detection approach is followed. The proposed algorithm reduces the search paths by performing iteratively a breadth-first tree search, which selects only K paths depending on the least metric. Note that in K-best detection, K denotes the number of selected paths and does not represent the number of nodes. The K-best detection reduces the complexity by reducing the number of search paths and facilitates the parallelization of computation by fixing the number of search paths. However, the limited coverage of the search tree results in an error floor and increases the BER. Moreover, for incorporating higher overloading, the number of paths needs to be increased which results in higher complexity.\par 
\subsection{Greedy algorithms}
The main disadvantage of the convex optimization and sphere decoding based algorithms is the higher complexity, which restricts its implementation in mMTC environment. To reduce the complexity of MUD, greedy algorithms were introduced for CS-MUD.
\subsubsection{Orthogonal least square and orthogonal matching pursuit}
To separate the nodes at the receiver exploiting the sparsity at symbol level, in \cite{rx_ga} orthogonal matching pursuit (OMP) and orthogonal least squares (OLS) are used for MUD, which iteratively select the most probable active users and subsequently estimate their data. For both OMP and OLS, the spreading influence, $\mathbf{S}$, in the sensing matrix, $\mathbf{{A}}$, is known at the base station, however, the influence  of channel, $\mathbf{H}$,  can be estimated. At each iteration of OMP, the column of sensing matrix, $\mathbf{A}$, which has higher correlation with the received signal is selected and the corresponding user is detected as active. In case of OLS, the selection is based on the minimum least square distance instead of correlations. OLS is more robust to errors than OMP at the cost of relatively higher complexity. However, as the greedy algorithms iteratively estimates the activity and data, it may suffer from error propagation. \par 
\subsubsection{Block-OLS and group-OMP}
The activity detection of CS-MUD using greedy algorithms was improved by exploiting the fact that when a node is active it transmits several bits in the current frame, therefore, makes the multiuser signal frame as block sparse \cite{rx_o1} \cite{rx_bols}. In \cite{rx_bols}, the OLS is extended to block-wise orthogonal least squares (BOLS) algorithm which detects the active user for a block of transmitted symbols. The activity is detected based on the sum of the minimum Euclidean distance of the spreading sequences from the block of $N_d$ symbols of the received signal. In the group orthogonal matching pursuit (GOMP) for CS-MUD \cite{rx_o1} instead of considering only the maximum correlated column of the spreading matrix with the received signal, the sum of the correlations of the $N_d$ received signals with the spreading sequence is set as the selection criteria which improves the activity detection. In \cite{rx_o1} to avoid the complexity of matrix inversion for larger $N_d$, the frame is divided into sub-frames and parallel detectors are deployed at the receiver to detect the signal. Furthermore, it is shown that false alarms, which are less critical as compared to miss detections, are reduced by incorporating activity-aware Viterbi decoder, which acts as a decision device for the activity of user on frame level. The activity detection was further improved by deploying a weighted GOMP (wGOMP) algorthim in \cite{rx_gait}. In  wGOMP algorithm, the weights are generated by the channel decoder  as \cite{rx_gait}
\begin{equation}
w_j=0.5+\frac{\xi_C(\hat{\textbf{d}}_k)-\xi_\epsilon (\hat{\textbf{d}}_k)}{2N_d} \:\:\:\:\:\: \forall j\in \Gamma(k),
\end{equation}
where $\xi_C(\hat{\textbf{d}}_k)$ is the Euclidean distance of the most likely codeword, the term $\xi_\epsilon(\hat{\textbf{d}}_k)=\| \hat {\textbf{d}}_k \|_2$ gives an indication about the activity of node $k$.  ${ \Gamma(k)}$ contains the vector indices corresponding to group $k$ and $N_d$ is the length of the  vector $\hat {\mathbf{d}}$. The activity detection is enhanced by multiplying these weights with the correlation in the group selection step of GOMP.  The symbol error rate (SER) of the CS-MUD by deploying wGOMP improved by magnitude of one as compared to GOMP and meet the oracle least square performance at SNR=20 dB \cite{rx_gait}.\par
In \cite{rx_scp},  MMV-CS model was considered and  simultaneous orthogonal matching pursuit algorithm (SOMP) is proposed. The algorithm is similar to GOMP algorithm and detects the support for a group of symbols. It is assumed that the sparsity remains the same for the whole frame.

\subsubsection{ Iterative order recursive least square}
The limitation of GOMP is that its complexity increases exponentially with the group size.  To reduce the higher complexity associated with the frame length, the iterative order recursive least square (IORLS) algorithm is proposed in \cite{rx_iorls}. IORLS iteratively employs the OMP algorithm and therefore, avoids the computations of group correlations.  The OMP node selection criteria is also enhanced by multiplying a weight matrix, $\mathbf{W}$, with the selection metric, i.e., the correlations of the sequences with the received signal. The weight matrix is a diagonal matrix where each entry, $\mathbf{W}_{n,n}$ represents the number of symbols for which the $n$-th sequence is detected as active  in the previous iteration, $1<n<N$ (number of sequences).  Moreover, the matrix inversion in OMP is replaced by order recursive least square to further reduce the complexity. The complexity of IORLS increases linearly with the number of iterations while that of GOMP increases exponentially with the number of symbols in a sub-frame. The performance of IORLS is dependent on the length of the frame and the number of iterations. \par
\subsubsection{Structured matching pursuit}
In the algorithms, which exploit block sparsity, the general assumption is that a node is active for several consecutive symbols. However, in mMTC some nodes may be active for several consecutive symbols while others may be active for a less number of symbols. Therefore, the active nodes set is changing within the duration of a frame. This structured sparsity in multiuser signal is considered in   \cite{rx_dn1} \cite{rx_dn2}   to improve the MUD. It is assumed that a portion of the active users remain active for several continuous time slots (common active users) while others changes at each time slot (dynamic active users). In \cite{rx_dn1}, it is assumed that the number of common active users is known a-prior. An algorithm called structured matching pursuit (SMP) algorithm is proposed for MUD. In SMP,  the common active users are detected first  based on the sum of their energies over the specified continuous  time slots. After the common active users are detected, the residual (initialized to the received signal)  is updated by subtracting the effect of the common active users. The dynamic active users are then detected at each time slot using OMP like selection criteria. The algorithm is compared with OMP, which detects the activity on symbol-by-symbol base, and it is shown that the performance is improved by magnitude of one at SNR 3 dB \cite{rx_dn1}. In \cite{rx_dn2} the idea was extended to estimate the data efficiently by exploiting this temporal correlation in the user activity. In the proposed dynamic compressive based MUD algorithm in \cite{rx_dn2} , the active users are detected at  symbol level and the support obtained for one symbol is used as an initial support for estimating the data at the next time slot. The proposed algorithm significantly improves the data estimation as compared to the conventional symbol-by-symbol OMP data estimation. \par  
\subsubsection{Matrix matching pursuit}
Considering the MMV-CS model, besides the MAP based algorithm in Section \ref{mpb}, the authors in \cite{rx_mmv} also presented a greedy algorithm called matrix matching pursuit (MMP) for the activity detection. MMP is an extension of OMP, which selects the active users based on the maximum correlation of the spreading sequences with the sample covariance matrix, $\Phi_{YY}$, of the received matrix $\mathbf{Y}$ in Equation (\ref{mmpe}).  The proposed algorithm improves the activity detection with the complexity invariant to the length of the frame. However, in \cite{rx_mmv}, the proposed algorithms are only for activity detection and a separate data detection has to be implemented once the support is obtained. Moreover, the performance evaluation considers only the AWGN channel and the effect of fading channels is not incorporated. Some parameters considered for the evaluation are not realistic in the context of mMTC, e.g., mMTC transmissions are of a few bytes while here a frame of 1000 bits is considered. The activity probability for mMTC is $p_a<0.1$ while in \cite{rx_mmv}, $p_a=0.35$ is considered for evaluating the effect of frame size.\par 
A summary of the CS-MUD algorithms is given in Table  \ref{mud}.

\begin{table*}[]
	\centering
	\caption{Comparison of CS-MUD Algorithms}
	\label{mud}
	\begin{tabular}{|c|c|c|c|c|}
		\hline
		\textbf{	Category } & \textbf{Scheme} & \textbf{Advantages} &\textbf{ Disadvantages}&\textbf{ Ref.} \\ \hline	
		\multirow{4}{*}{ \vspace{-7cm} 
			\begin{minipage}[t]{0.15\textwidth}
				MAP based Algorithms
		\end{minipage} } 
		& 
		\begin{minipage}[t]{0.22\textwidth}
			\vspace{0.1cm}
			S-MAP\\ Algorithms 
		\end{minipage} 	      & 
		\hspace{-0.45cm}\begin{minipage}[t]{0.35\textwidth}
			\begin{itemize}
				\vspace{0.1cm}
				\item Exploits sparsity to detect the user activity and  avoid control signaling overhead 
				\vspace{0.18cm}
				\item Robust to asynchronous transmissions  
				\vspace{0.1cm}
			\end{itemize}
		\end{minipage}       & 	
		\hspace{-0.45cm}\begin{minipage}[t]{0.35\textwidth}
			\vspace{0cm}
			\begin{itemize}
				\item Higher complexity
				\vspace{0.2cm}
				\item Not focused on overloaded systems
			\end{itemize}
		\end{minipage}   &   \begin{minipage}[t]{0.08\textwidth}
			\vspace{00.2cm}
			\cite{e1}   
		\end{minipage}   \\ \cline{2-5}  & 
		\begin{minipage}[t]{0.22\textwidth}
			\vspace{0.1cm}
			Approximate MAP  Algorithm \\
			(\footnotesize{MMV-CS}) 
		\end{minipage} 	      & 
		\hspace{-0.45cm}\begin{minipage}[t]{0.35\textwidth}
			\begin{itemize}
				\vspace{0.1cm}
				\item Complexity is independent  of frame length
				\vspace{0.18cm}
				\item Robust to asynchronous transmissions  
				\vspace{0.15cm}
			\end{itemize}
		\end{minipage}       & 	
		\hspace{-0.45cm}\begin{minipage}[t]{0.35\textwidth}
			\vspace{0cm}
			\begin{itemize}
				\item Additional data estimation is required
				\vspace{0.1cm}
				\item High complexity than greedy algorithms		\end{itemize}
			\vspace{0.1cm}
		\end{minipage}   &   \begin{minipage}[t]{0.08\textwidth}
			\vspace{00.2cm}
			\cite{rx_mmv}   
		\end{minipage}   \\ \cline{2-5}  & 
		\begin{minipage}[t]{0.22\textwidth}
			\vspace{0.20cm}
			Sphere decoding 
		\end{minipage} 	      & 
		\hspace{-0.45cm}\begin{minipage}[t]{0.25\textwidth}
			\begin{itemize}
				\vspace{0.2cm}
				\item Maximum a posteriori performance
			\end{itemize}
		\end{minipage}       & 	
		\hspace{-0.45cm}\begin{minipage}[t]{0.35\textwidth}
			\vspace{0cm}
			\begin{itemize}
				\item No guarantee to terminate in polynomial time
				\vspace{0.15cm}
				\item No possibility to parallelize the computations
			\end{itemize}
			\vspace{0.1cm}
		\end{minipage}   &   \begin{minipage}[t]{0.08\textwidth}
			\vspace{00.2cm}
			\hspace{-0.02cm}\cite{rx_sd1,rx_sd2,rx_np}   
		\end{minipage}   \\ \cline{2-5} 	&
		\begin{minipage}[t]{0.22\textwidth}
			\vspace{0.1cm}
			K-best Detection for sphere decoding 
		\end{minipage}         &   	
		
		\hspace{-0.55cm}	\begin{minipage}[t]{0.35\textwidth}
			\begin{itemize}
				\vspace{0cm}
				\item  Constant run time 
				\vspace{0.15cm}
				\item Allow parallelization and pipelining
			\end{itemize}
		\end{minipage}            & 
		
		\hspace{-0.55cm} \begin{minipage}[t]{0.35\textwidth}
			\vspace{0cm}
			\begin{itemize}
				\item  Complexity increases with overloading the system
				\vspace{0.1cm}
				\item  BER floor due to limited search paths
			\end{itemize}
			\vspace{0.1cm}
		\end{minipage}                        &   \begin{minipage}[t]{0.08\textwidth}
			\vspace{00.2cm}
			\cite{rx_kb}
		\end{minipage}        \\  \hline
		\multirow{6}{*}{\vspace{-7cm} 
			\begin{minipage}[t]{0.15\textwidth}
				Greedy \\ Algorithms
		\end{minipage}}
		
		& 
		\begin{minipage}[t]{0.22\textwidth}
			\vspace{0.12cm}
			OMP    
		\end{minipage}      &    \hspace{-0.55cm} \begin{minipage}[t]{0.35\textwidth}
			\vspace{0.1cm}
			\begin{itemize}
				\item Lower complexity as compared to other greedy algorithms, e.g, OLS
			\end{itemize}
			\vspace{0.1cm}
		\end{minipage}   &      
		\hspace{-0.55cm} \begin{minipage}[t]{0.35\textwidth}
			\begin{itemize}
				\vspace{0.1cm}
				\item Relatively higher BER
			\end{itemize}
		\end{minipage}      &      \vspace{0cm}   \begin{minipage}[t]{0.08\textwidth}
			\vspace{0cm}
			\cite{rx_ga}  
		\end{minipage}  \\ \cline{2-5} 
		& 
		\begin{minipage}[t]{0.22\textwidth}
			\vspace{0cm}
			OLS    
		\end{minipage}                      &    
		
		\hspace{-0.55cm} \begin{minipage}[t]{0.35\textwidth}
			\begin{itemize}
				\vspace{0.0cm}
				\item Lower BER than OMP
			\end{itemize}
			\vspace{0.1cm}
		\end{minipage}    &        
		\hspace{-0.55cm} \begin{minipage}[t]{0.35\textwidth}
			\begin{itemize}
				\vspace{0cm}
				\item Higher complexity than OMP
				\vspace{0.12cm}
			\end{itemize}
		\end{minipage}            &        \begin{minipage}[t]{0.08\textwidth}
			\vspace{-0.1cm}
			\hspace{-0.2cm}
			\cite{rx_ga,rx_gait}    
		\end{minipage}   \\ \cline{2-5} 
		& 
		
		\begin{minipage}[t]{0.22\textwidth}
			\vspace{0.3cm}
			GOMP  
		\end{minipage}                     &   
		\hspace{-0.55cm} \begin{minipage}[t]{0.35\textwidth}
			\begin{itemize}
				\vspace{0cm}
				\item  Exploits block sparsity
				\vspace{0.1cm}
				\item Higher activity detection accuracy 
			\end{itemize}
		\end{minipage}     &
		\hspace{-0.55cm} \begin{minipage}[t]{0.35\textwidth}
			\vspace{0.0cm}
			\begin{itemize}
				\item Complexity increases exponentially with group size
				\vspace{0.15cm}
				\item Performance gain depends on the frame size
			\end{itemize}
			\vspace{0.05cm}
		\end{minipage}               &            \begin{minipage}[t]{0.08\textwidth}
			\vspace{-0.1cm}
			\hspace{-0.15cm}
			\cite{rx_o1,rx_bols,rx_dn1}  
		\end{minipage}   \\ \cline{2-5} 	
		&   \begin{minipage}[t]{0.22\textwidth}
			\vspace{0.16cm}
			IORLS    
		\end{minipage}                   &        
		\hspace{-0.55cm} \begin{minipage}[t]{0.35\textwidth}
			\vspace{0.0cm}
			\begin{itemize}
				\item  Exploit block sparsity
				\vspace{0.1cm}
				\item No matrix inversion
				\vspace{0.1cm} 
				\item Robust to noise
			\end{itemize}
			\vspace{0.1cm}
		\end{minipage} 	    &
		\hspace{-0.55cm} \begin{minipage}[t]{0.35\textwidth}
			\begin{itemize}
				\vspace{0cm}
				\item The performance gain comes from large frame size, while in mMTC data packet is typically of small size. 
				\vspace{0.15cm}
			\end{itemize}
		\end{minipage}                     &     \begin{minipage}[t]{0.08\textwidth}
			\vspace{0.1cm}
			\cite{rx_iorls}  
		\end{minipage}     \\ \cline{2-5} 
		
		& 
		
		\begin{minipage}[t]{0.22\textwidth}
			\vspace{0cm}
			SOMP  \\
			\scriptsize{(MMV-CS)}
		\end{minipage} 	      & 
		\hspace{-0.45cm}\begin{minipage}[t]{0.35\textwidth}
			\vspace{0.0cm}
			\begin{itemize}
				\item Memory reduction 
				\vspace{0.1cm}
				\item Faster detection
				\vspace{0.1cm}
				\item Scalable
			\end{itemize}
			\vspace{0.1cm}
		\end{minipage}       & 	
		\hspace{-0.45cm}\begin{minipage}[t]{0.35\textwidth}
			\begin{itemize}
				\vspace{0cm}
				\item Computational complexity increases with measurement vectors
				\vspace{0.05cm}
			\end{itemize}
		\end{minipage}   &     \begin{minipage}[t]{0.08\textwidth}
			\vspace{0.1cm}
			\cite{rx_scp}  
		\end{minipage}    \\ \cline{2-5} 	&
		
		\begin{minipage}[t]{0.22\textwidth}
			\vspace{0.12cm}
			MMP \\
			\scriptsize{(MMV-CS)}
		\end{minipage}         &   	
		
		\hspace{-0.55cm}	\begin{minipage}[t]{0.35\textwidth}
			\vspace{0.0cm}
			\begin{itemize}
				\item Complexity is independent  of Frame length
				\vspace{0.1cm}
				\item Robust to asynchronous transmissions
			\end{itemize}
			\vspace{0.1cm}
		\end{minipage}            & 
		
		\hspace{-0.55cm} \begin{minipage}[t]{0.35\textwidth}
			\begin{itemize}
				\vspace{0cm}
				\item Additional data estimation step
				\vspace{0.1cm}
				\item Simulation parameters are not realistic for mMTC 
			\end{itemize}
			\vspace{0.05cm}
		\end{minipage}                        &       \begin{minipage}[t]{0.08\textwidth}
			\vspace{0.1cm}
			\cite{rx_mmv}  
		\end{minipage}  \\ \cline{2-5} \hline
	\end{tabular} 
\end{table*}

\section{CS based medium access}\label{mad}
In mMTC, the small data packet size necessitates the use of request grant free medium access scheme. CS-MUD is introduced to facilitate a grant free multiple access scheme in the physical layer. To enhance the medium access scheme, many solutions are proposed in the literature, which can be categorized into the following categories.
\subsection{Baseline non-orthogonal medium access}
Most of the current research considers the conventional CDMA-like medium access in the physical layer, where a dedicated spreading code is assigned to each node in time domain. These code sequences act as signatures to distinguish the active nodes. The data frame of the node consists of only the payload and no extra control signaling is needed for medium access. As every node has dedicated sequence, there is no collision in accessing the medium access, however, the sequences are non-orthogonal and have some correlation. Authors in \cite{sp_mc1,sp_mc2,sp_mc3} improved the medium access scheme by introducing the CS-MUD in multicarrier CDMA which is named as multicarrier compressed sensing based multiuser detection (MCSM). The proposed scheme improves the scalability and flexibility of accessing both the time and frequency resources. The spectral efficiency of the system can be improved by reducing the number of subcarriers per node.
To gain frequency diversity in MCSM, a scheduling technique for allocating the subcarriers is introduced in \cite{sp_mc2}. In the proposed scheduling scheme, the set of allocated subcarriers changes after a predefined number of symbols. Furthermore, the subcarriers are allocated such that they lie within the coherence bandwidth of the channel, which facilitates the use of non-coherent modulation and avoids pilot transmissions for channel estimation. A hardware implementation is also demonstrated in \cite{sp_mc3}.\par 
Although, the non-orthogonal medium access scheme with CS-MUD facilitates the grant free medium access and increases the spectral efficiency, the main challenge is the activity detection in massive mMTC. In CS-MUD, the activity detection is dependent on the correlation between the spreading sequences. For a fixed length of the spreading sequence, increasing the number of sequences increases the correlation between sequences. Therefore, in case of massive mMTC where the number of devices is higher, allocating low correlated sequences to each node will be a challenging task.       
\subsection{Enhanced non-orthogonal medium access}
\begin{figure}[!h]
	\centering
	\includegraphics[scale=.7]{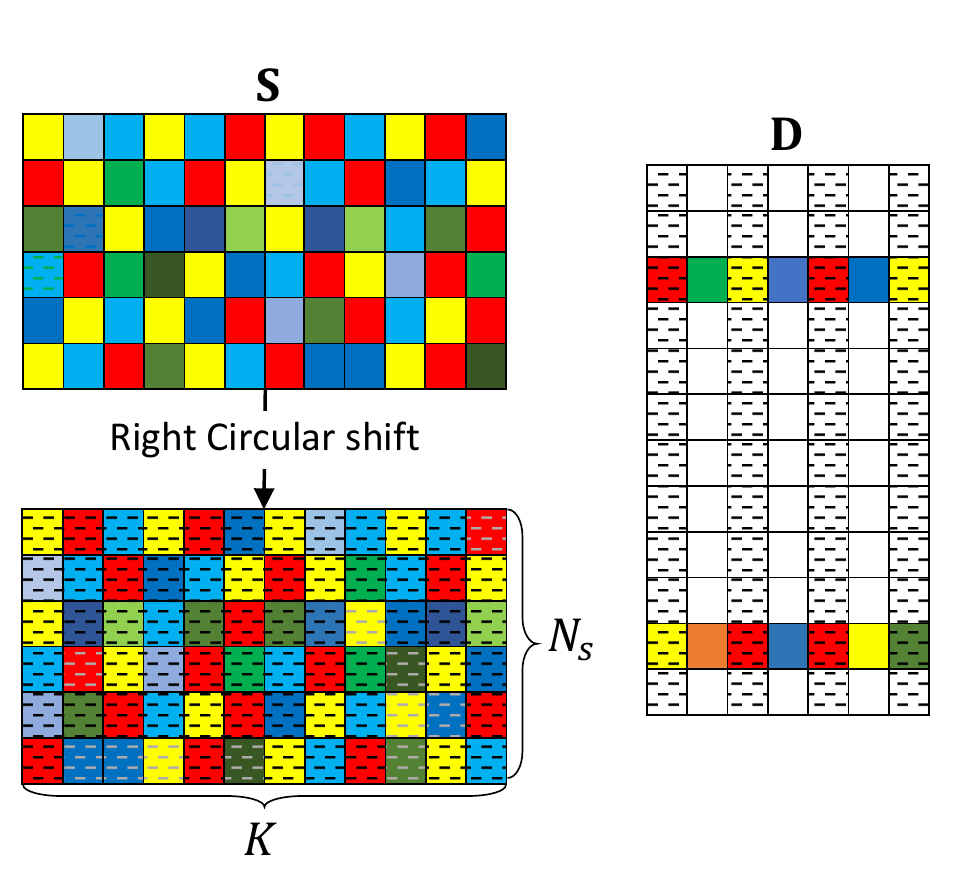}
	\caption {Matrix representation of data spreading:  \textit{even} indexed columns of $\mathbf{D}$  are spread over matrix $\mathbf{S}$ and  \textit{odd} indexed columns of $\mathbf{D}$ over shifted $\mathbf{S}$ }
	\label{sp}
\end{figure}

The medium access scheme in the MCSM is enhanced in \cite{sp_oe} by introducing spreading sequence diversity. Each user uses two spreading sequences from the sensing matrix, $\mathbf{A}$, to spread their data  frame $\mathbf{d}$. Half of the data symbols are spread over one sequence and the other half over the other as shown in Figure \ref{sp}.
At the receiver, for MUD the correlation between the received symbols and the spreading sequence is now averaged over two sequences, which results in more accurate activity detection. Due to this spreading diversity, the MUD in a non-orthogonal multiple access scheme is improved for a given number of users. In other words, more number of users can be accommodated while maintaining the same BER, as compared to the single sequence spreading.   \par
Although, the proposed enhanced non-orthogonal medium access scheme improves the performance and makes the possibility of increasing the overloading, it still has dependency on the correlations of the spreading sequences for accurate activity detection.   
\subsection{Multiple sequence non-orthogonal medium access}
\begin{figure}[!h]
	\centering
	\includegraphics[scale=.75]{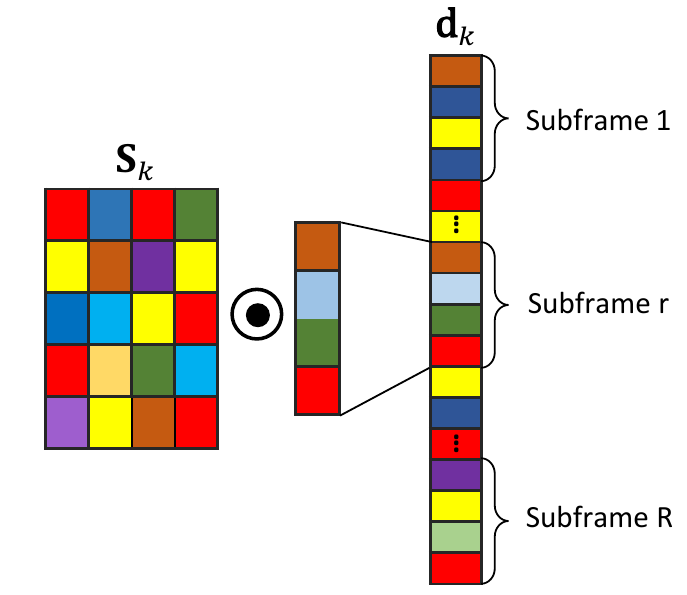}
	\caption {Grouping of data symbols and spreading over multiple sequences: $\odot
		$ represent the spreading of first symbol over first column of $\mathbf{S}_k$, second symbol over second column and so on}
	\label{ms12}
\end{figure}

In \cite{sp_ms1} \cite{sp_ms2} the problem of scarcity of low correlated spreading sequences for
massive number of nodes is addressed and multiple sequence based medium access is proposed. In the proposed scheme, the data frame of the $k$-th user, $\mathbf{{x}}_k$,  is divided into $R=N_d/v$ sub-frames, where $v$ is the number of symbols in one subframe such that $N_d$ mod $v=0$. Instead of allocating a dedicated single  sequence, sets of sequences are generated from a pool of sequences.  Each user randomly selects a set of sequences, $\mathbf{S}_k \in \mathbb{C}^{N_s\times \nu}$, from the pool of sequence sets. These sequence sets are used to spread the  subframes which  are then transmitted as depicted in Figure \ref{ms12}. 
The activity detection in the multiple sequence based transmission improves due to the averaging of the correlations between the sequences as compared to a single sequence. As the medium is randomly accessed and there is no dedicated sequence, the data frames are also embedded with the frame IDs, which are used to identify the active nodes. The multiple sequence scheme improves the activity detection and limits the problem of scarcity of sequences by using random   access.  However, as the users select the sequences randomly there is a probability of collision and increasing the number of sequence sets increases the correlations, which consequently increases the probability of detection errors. Moreover, the large number of sequence sets also increases the complexity of the multiuser detector. 
\subsection{Codebook based non-orthogonal medium access}
\begin{figure}[!h]
	\centering
	\includegraphics[scale=.65]{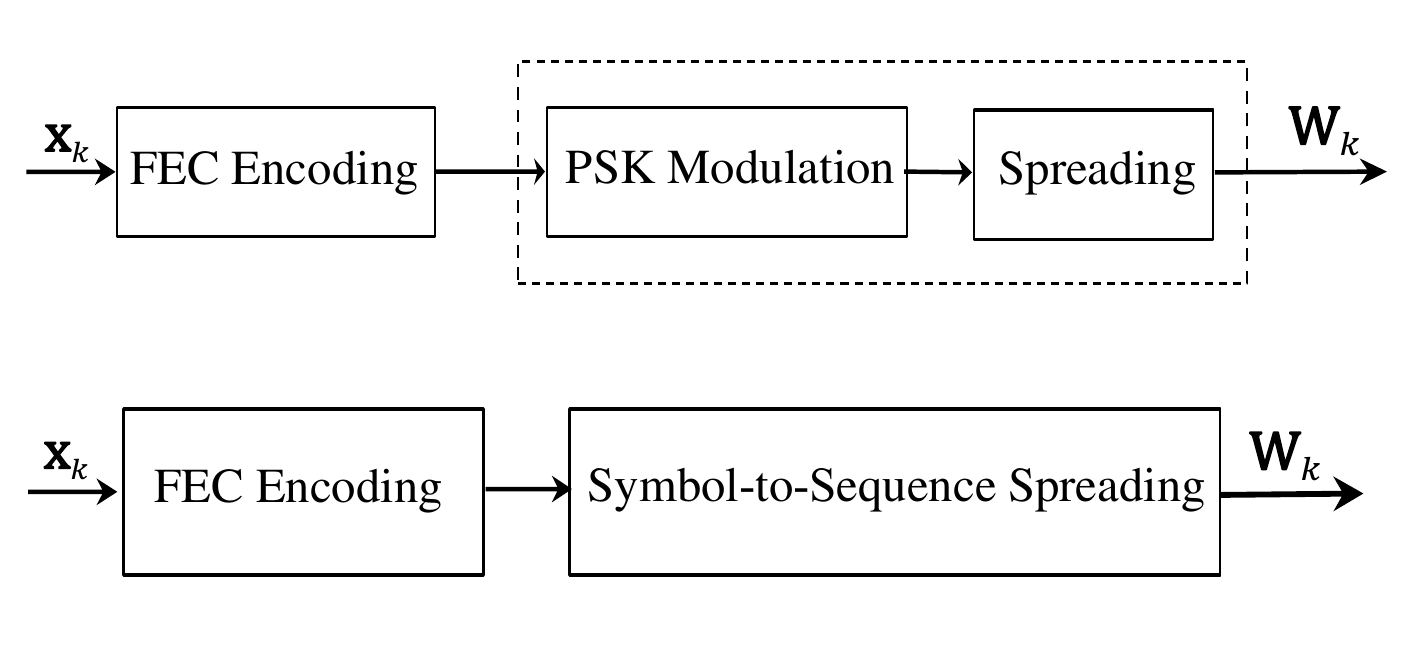}
	\caption {Direct symbol-to-sequence spreader}
	\label{en}
\end{figure}

In \cite{sp_cb} a codebook based non-orthogonal CDMA scheme is proposed. The modulation and spreading processes are merged together into a direct symbol-to-sequence spreader as shown in Figure \ref{en}.  

Each user is assigned with a unique codebook and the incoming bits are directly mapped to a codeword in the codebook. The codewords are the spreading sequences, which are designed such that they have low correlations. The spreading sequences in the codebook are considered as codewords drawn from a  multidimensional  constellation, therefore, for higher modulation scheme the Euclidean distance between the codewords doesn’t decrease much as compared to the two dimensional constellation. Therefore, the required SNR at higher modulation schemes (i.e., modulation order greater than 4) is less as compared to that of the scheme, which involves PSK modulation. \par 
The codebook based CDMA with CS-MUD simplifies the encoding process at the sensor nodes, which results in long battery life and increases the spectral efficiency at the cost of minimum performance loss. However, the proposed scheme is sensitive to multiple access interference and a more robust decoding scheme is required for situation where the number of active users is high.
\subsection{Channel division medium access}
\begin{figure}[!h]
	\centering
	\includegraphics[scale=0.7]{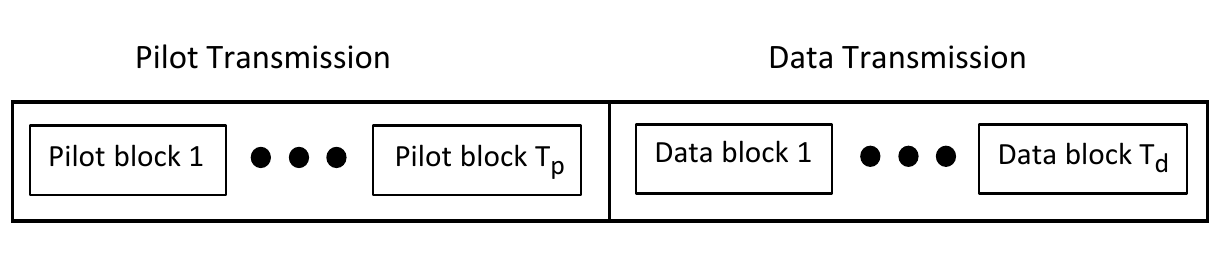}
	\caption {Frame structure of channel division medium access transmission}
	\label{fs}
\end{figure}

The increase of correlation between the sequences with the increase in the number of devices limits the performance of the NOMA, which uses spreading sequences as the signature. Moreover, for situation where there are massive number of nodes, it would be a challenging task to assign dedicated sequences to each node.  In \cite{sp_ch1}, a channel division multiple access scheme is proposed which uses the channel state information (CSI) as signature for separating the active users at the receiver. The active users transmits pilot sequences, which are randomly chosen from a set of sequences. At the receiver, the CSI is estimated and consequently the active users are detected based on their respective CSI. As the pilot sequences are randomly selected, there is a probability of pilot sequence collision. The effect of pilot collision is reduced by sending multiple pilot sequences within a frame. The frame structure of the proposed scheme is depicted in Figure \ref{fs}, where each frame of a node consists of $T_p$ pilot blocks and $T_d$ data blocks. \par 

This approach allows the system to overload without the limitation of the number of spreading sequences. However, to separate the users’ data patterns, the CSI of users should be highly uncorrelated which means that the channel impulse response should be of larger size. The proposed scheme will not work for a flat fading channel because it would not be possible to differentiate the CSI of different users.\par 
A similar scheme is also adopted in \cite{o_p1}, in which the number of possible Zadoff-chu and power residue based pilot sequences are calculated. For Zadoff-chu the possible number of sequences in sequence set is $ P = (N_s-1)(N_s/V_t   +1)$, where $N_s$ is length of sequence, and $V_t$ is the smallest integer greater than the number of channel taps. For power residue based, $P = (R-1)N_s$ ,  where $R$ is a positive integer which divides $N_s-1$. \par 
A summary of the CS based medium access schemes is presented in Table \ref{spt}.  
\begin{table*}[]
	\centering
	\caption{Comparison of medium access schemes}
	\label{spt}
	\begin{tabular}{|c|c|c|c|c|}
		\hline
		\textbf{	Scheme  }                         & \textbf{Signature   }                       & \textbf{Frame structure}        & \textbf{Medium access                             }                                         & \textbf{Reference }\\ \hline
		
		\begin{minipage}[t]{0.23\textwidth}
			\vspace{0cm}
			Baseline Non-orthogonal medium access   
			\vspace{0.1cm}	
		\end{minipage}	
		&     
		\begin{minipage}[t]{0.20\textwidth}
			\vspace{0cm}
			Node specific sequence of the sensing matrix 
			\vspace{0.1cm}	
		\end{minipage}
		
		&   
		\begin{minipage}[t]{0.18\textwidth}
			\vspace{0cm}
			Only data block 
			\vspace{0.1cm}	
		\end{minipage}
		& 
		\begin{minipage}[t]{0.15\textwidth}
			\vspace{0cm}
			Direct access 
			\vspace{0.1cm}	
		\end{minipage}
		
		&    
		
		\begin{minipage}[t]{0.09\textwidth}
			\vspace{0cm}
						\hspace{-0.2cm}
			\cite{sp_mc1,sp_mc2,sp_mc3}   
		\end{minipage}

		\\ \hline
		\begin{minipage}[t]{0.23\textwidth}
			\vspace{0cm}
			Enhanced Non-orthogonal medium access 
			\vspace{0.1cm}
		\end{minipage}  			
		&

		\begin{minipage}[t]{0.20\textwidth}
			\vspace{0cm}
			Two sequences from the same sensing matrix  
			\vspace{0.1cm}
		\end{minipage}  
		& 
		
		\begin{minipage}[t]{0.18\textwidth}
			\vspace{0cm}
			Only data block   
			\vspace{0.1cm}
		\end{minipage}  
		& 
		\begin{minipage}[t]{0.15\textwidth}
			\vspace{0cm}
			Direct access 
			\vspace{0.1cm}
		\end{minipage} 
		
		&       
		
		\begin{minipage}[t]{0.09\textwidth}
			\vspace{0.1cm} \cite{sp_oe} 
			\vspace{0.1cm}
		\end{minipage} 
		
		\\ \hline

		\begin{minipage}[t]{0.23\textwidth}
			\vspace{0cm}
			Multiple sequence based 	Non-orthogonal medium access  
			\vspace{0.1cm}
		\end{minipage}  		
		&                  
		
		\begin{minipage}[t]{0.18\textwidth}
			\vspace{0cm}
			Frame ID     
			\vspace{0.1cm}
		\end{minipage} 
		
		&
		
		\begin{minipage}[t]{0.18\textwidth}
			\vspace{0cm}
			Frame ID + data block  
			\vspace{0.1cm}
		\end{minipage}

		& 
		\begin{minipage}[t]{0.15\textwidth}
			\vspace{0cm}
			Random access 
			\vspace{0.1cm}
		\end{minipage}    
		
		&         
		
		\begin{minipage}[t]{0.09\textwidth}
			\vspace{0cm}
		\hspace{-0.28cm}
			\cite{sp_ms1}, \cite{sp_ms2}
			\vspace{0.1cm}
		\end{minipage}  
		
		\\ \hline
		\begin{minipage}[t]{0.23\textwidth}
			\vspace{0cm}
			Codebook based 	Non-orthogonal medium access  
			\vspace{0.1cm}
		\end{minipage}  
		& 
		\begin{minipage}[t]{0.20\textwidth}
			\vspace{0cm}
			Codebook     
			\vspace{0.1cm}
		\end{minipage}      & 
		
		\begin{minipage}[t]{0.18\textwidth}
			\vspace{0cm}  
			Only data block  
			\vspace{0.1cm} 
		\end{minipage}                 &  
		\begin{minipage}[t]{0.15\textwidth}
			\vspace{0cm}
			Direct access 
			\vspace{0.1cm}
		\end{minipage}           &   		   
		\begin{minipage}[t]{0.09\textwidth}
			\vspace{0cm}
			\cite{sp_cb}
			\vspace{0.1cm}
		\end{minipage} 	                           \\ \hline
		
		\begin{minipage}[t]{0.23\textwidth}
			\vspace{0cm}  
			Channel division random access  
			\vspace{0.15cm} 
		\end{minipage} 	
		&   
		
		\begin{minipage}[t]{0.20\textwidth}
			\vspace{0cm}  
			Channel state information    
			\vspace{0.15cm} 
		\end{minipage} 
		& 
		
		\begin{minipage}[t]{0.18\textwidth}
			\vspace{0cm}  
			Pilot block + data block
			\vspace{0.15cm} 
		\end{minipage}     
		& 
		\begin{minipage}[t]{0.15\textwidth}
			\vspace{0cm}  
			Random access
			\vspace{0.1cm} 
		\end{minipage}

		&     
		\begin{minipage}[t]{0.09\textwidth}
			\vspace{0cm}  
			\cite{sp_ch1,sp_ch2} 
			
			\vspace{0.15cm} 
		\end{minipage} 
		
		\\ \hline
	\end{tabular}
\end{table*}

\section{ Combination of CS-MUD with other techniques } \label{hyb}
The method of CS-MUD was combined with other different techniques to enhance the overall performance of the machine type communication system. \par
\begin{figure}[!h]
	\centering
	\includegraphics[scale=.53]{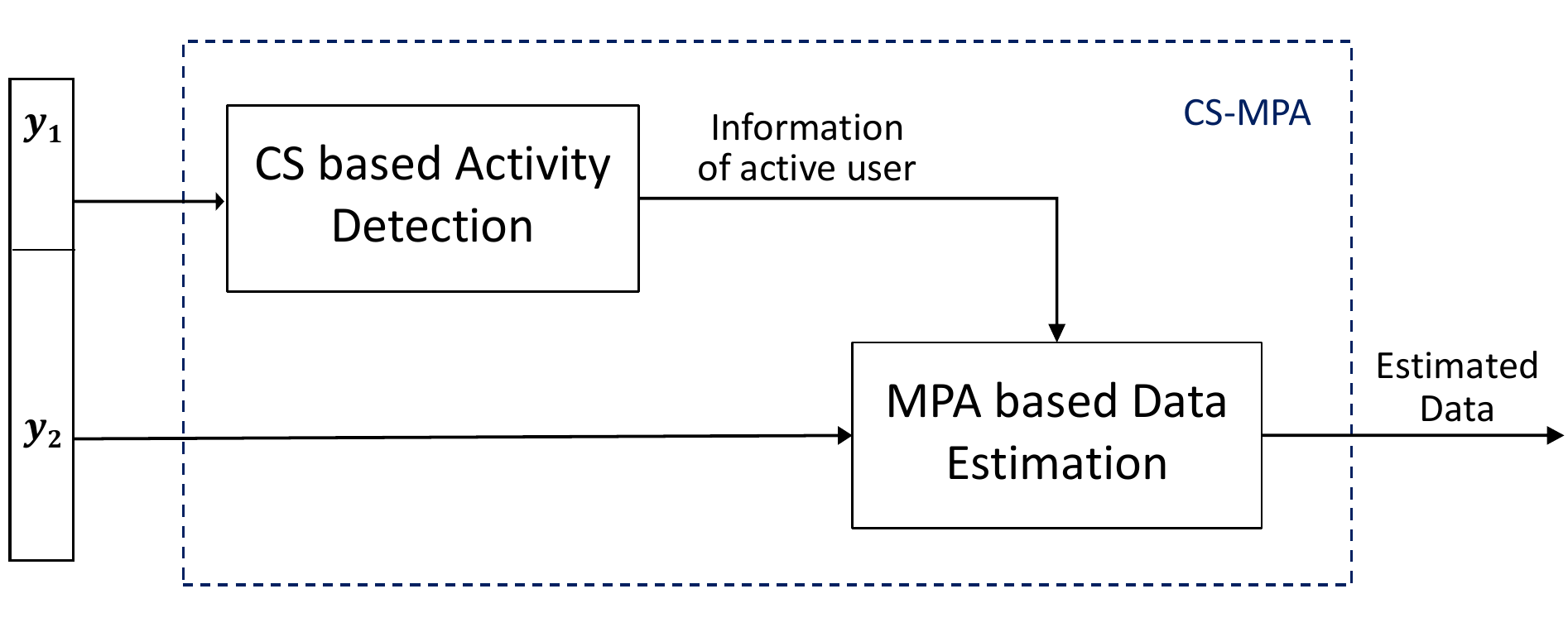}
	\caption {Compressive sensing based MPA receiver}
	\label{csmpa}
\end{figure}

 
The non-orthogonal low density signature based orthogonal frequency division multiplexing and CDMA (LDS-OFDM/CDMA) scheme exploits the sparse signatures, which allows the use of a relatively low complexity message passing algorithm (MPA) to approximate the MAP detection. However, the basic assumption in MPA is that the activity is known at the receiver, which require a higher control signaling overhead. In \cite{o_lds1}, CS-MUD is combined with MPA receiver and CS-MPA receiver is proposed to optimize the activity and data detection. An active node transmits a dedicated dense spreading sequence before transmitting the actual symbols spread over the sparse LDS signatures. The multiplexed data frame of all active nodes constitutes $\mathbf{y}_1$ and $\mathbf{y}_2$. $\mathbf{y}_1$ represents the superimposed dense signatures from all active nodes while $\mathbf{y}_2$  is the superimposed sparse LDS signatures over which the data symbols are spread. At the receiver the CS-MUD takes $\mathbf{y}_1$ as input and uses compressive sensing algorithm  to detect the active users  prior to  the MPA detector. The support obtained by the CS-MUD and the $\mathbf{{y}}_2$ are fed into the MPA to efficiently estimate the data. The detection process is depicted in Figure \ref{csmpa}.

When the activity is not known, the MPA has to search for all the possible combinations of data colliding at a subcarrier, which increases the complexity as well as degrades the performance.  In the  CS-MPA receiver when the support is fed into the MPA detector, the possible combinations for data estimation are reduced which enhances the data estimation accuracy as well as reduces the complexity. The CS-MPA is optimized in \cite{o_lds2} by incorporating a two stage CS based activity detection. At the first stage a correlation based activity detection is carried out. The approximated support obtained at the first stage is fed into the second stage which executes compressive sampling matching pursuit (CoSaMp) \cite{needell2009cosamp} algorithm. The CoSaMP uses the $\mathbf{{y}}_1$ and the estimated support in the first stage as prior information and detect the activity more accurately. \par
The LDS based non-orthogonal schemes are capable of accommodating more users on less physical resources when all users are active while the non-orthogonal CDMA with CS-MUD only works if a small number of the users are active. For example for 6 users sharing 4 physical resources, using the LDS based NOMA, when all 6 users are active, the MPA algorithm is capable of separating the data of all users. On the other hand the CS-MUD cannot separate users when the number of active users is greater than half of the number of physical resources. However, in mMTC where only small number of users are active, the LDS based NOMA requires control signaling overhead for medium access while non-orthogonal CDMA with CS-MUD does not need any control signaling. \par
Considering the variations in the sparsity level of the multiuser signal, in \cite{o_sw} a switching mechanism between the CS-MUD and the classical multiuser data detection is proposed. A threshold for the sparsity level based on the OMP algorithm is derived. The multiuser detector uses the OMP algorithm if the sparsity is high and is switched to the classical linear minimum mean square error estimator when the sparsity is lower than the threshold. \par  
To efficiently handle the massive connectivity, the MAC layer coded slotted ALOHA  \cite{aloha} and the PHY layer CS-MUD are combined in \cite{o_cra1} \cite{o_cra2}. The CS-MUD exploits the sparsity at physical layer to jointly detect the activity and data while the coded slotted ALOHA is capable of resolving the collisions by exploiting the successive interference cancellation. The joint activity and data detection at physical layer facilitates the efficient collision resolution and increases the user density at MAC layer. \par 
In \cite{o_mimo1}, a large-scale spatial modulation multiple-input-multiple-output system is proposed to reduce the radio frequency chains and hence reduce the hardware cost and power consumption. Each user selects one antenna element out of all the antenna elements used for data transmission. The MUD at the base station is posed as underdetermined signal detection problem and CS-MUD is used for MUD.  \par 
\section{Performance comparison of CSMUD schemes}\label {comparison}
\begin{figure*}[htp]
	\centering
	\subfigure[]{\includegraphics[scale=0.55]{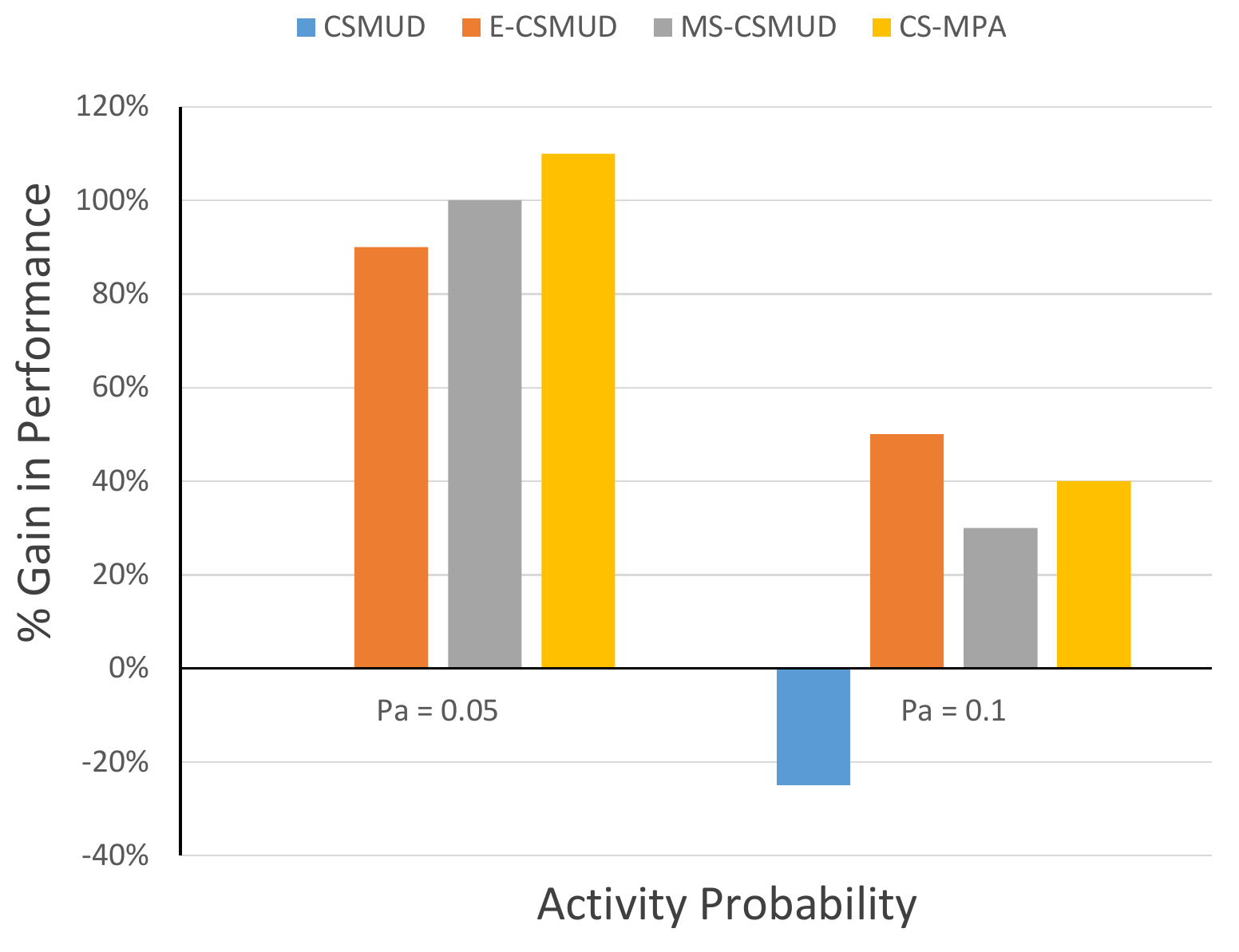}}\quad\quad
	\subfigure[]{\includegraphics[scale=0.55]{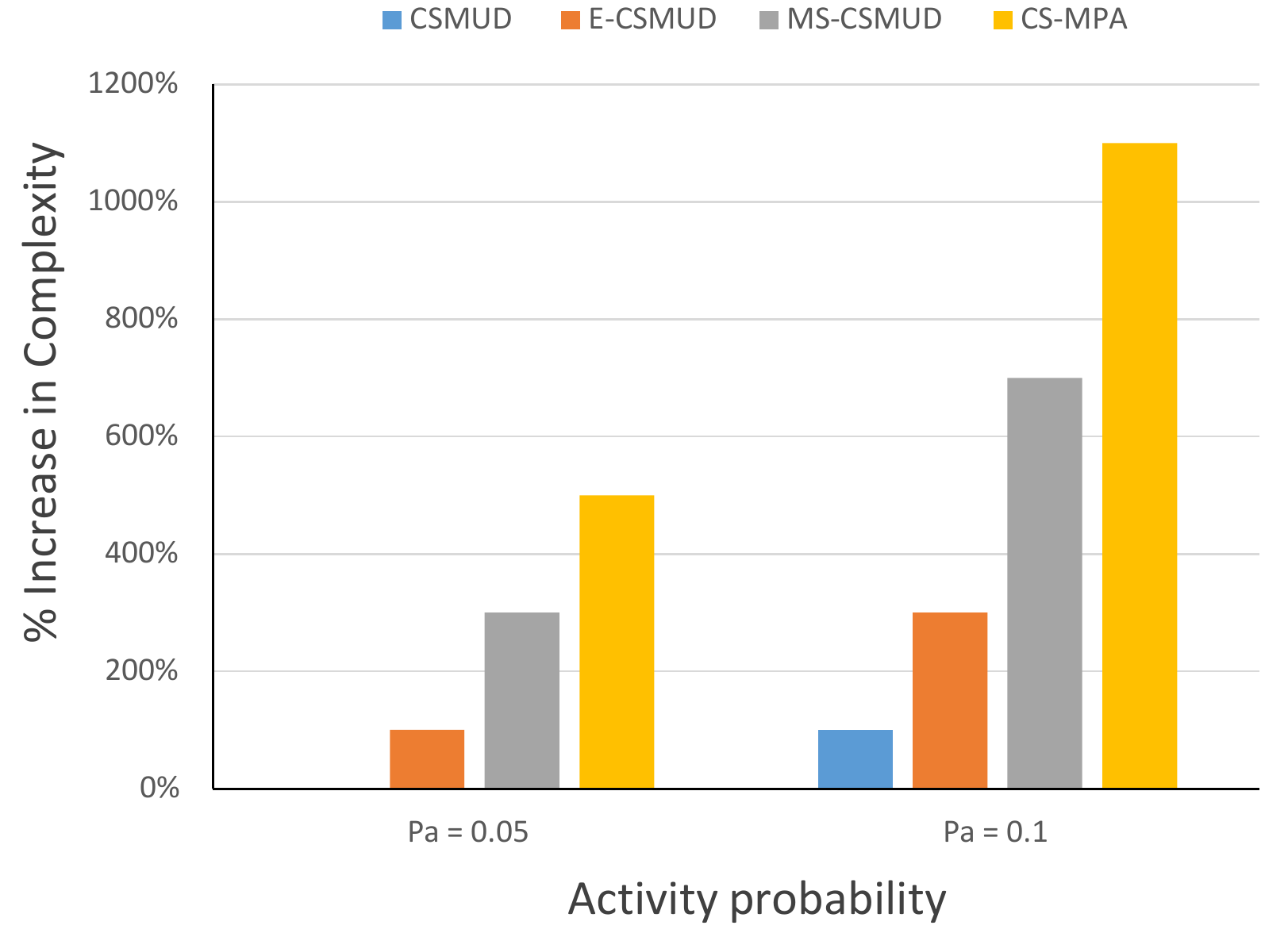}}
	
	\caption{\label{cmpr} Comparison of CS based MUD schemes in terms of (a) Performance and (b) Complexity }
\end{figure*}

The available CS based MUD schemes are compared in Figure \ref{cmpr}, taking the single sequence CSMUD at activity probability, $P_a = 0.05$, and overloading factor, $\lambda = 4$, as the baseline.  The effect of increasing the activity probability from $P_a=0.05$ to $P_a=0.1$ on the BER performance is shown for different CS based MUD schemes. For $P_a=0.05$, the performance of the enhanced-CSMUD (E-CSMUD), multisequence based CSMUD (MS-CSMUD) and CS based MPA (CS-MPA) is significantly better than the baseline CSMUD. The improvement in performance for E-CSMUD and MS-CSMUD comes from averaging the correlation over multiple sequences while for the CS-MPA it comes from the use of separate activity detection by GOMP and data detection by MPA. The MS-CSMUD performs slightly better than the E-CSMUD because of the multiple spreading sequences used by a single user.  When the activity probability is increased to $P_a=0.1$, a performance degradation is observed due to higher multiple access interference, however, the rate of performance degradation varies for  different schemes.  The baseline CSMUD suffers a performance loss of 25\% due to the increase in correlations between the sequences. The rate of performance degradation of  E-CSMUD is less than that of the baseline due to the averaging of correlation over two spreading sequences. The performance degradation rate in MS-CSMUD is higher than that of the E-CSMUD because in MS-CSMUD the users retrieve the spreading sequences randomly from a pool of sequences, therefore, with increase in the number of active nodes, the probability of collision increases, which increases the bit error rate in MS-CSMUD.  Moreover, in MS-CSMUD the users are identified by decoding the user IDs that are included within the data frame, the probability of wrong detection increases with increase in the activity probability. The CS-MPA performs well in low activity probability, however, for higher number of active nodes, two sources contributes in performance degradation. First, the correlation between spreading sequences increases and secondly, the multiuser interference at the non-zero chips of the spreading sequences increases. \par 
The performance improvement in the variants of CSMUD schemes comes at the cost of higher computation complexity. The dominant operation in CS decoding algorithms, e.g., GOMP, is the pseudoinverse of the sensing matrix. The E-CSMUD uses two sequences per user due to which the complexity of E-CSMUD is twice as that of the baseline. Similarly, for group size of four, the MS-CSMUD uses four sequences for each user, which makes its complexity four times as the baseline. Moreover, the decoding of the frame ID also contributes in increasing the complexity of MS-CSMUD. The CS-MPA is a two stage decoding technique with highest complexity amongst the CS based MUD schemes. Besides the complexity of GOMP, the major complexity in CS-MPA comes from the implementation of a separate message passing algorithm for data detection. The complexity of the MPA depends on the number of superimposed users over a single chip of the spreading sequence.

\section{Concluding Remarks} \label {con}
This survey article summarizes all the relevant algorithms and schemes of NOMA based on CS-MUD which is a promising solution for massive connectivity in mMTC. CS-MUD leverages the sporadic nature of mMTC and the theoretical foundation of compressive sensing techniques to provide an elegant way to tackle the NOMA problem. NOMA with CS-MUD significantly simplifies the encoding process and enables request-grant free medium access scheme for the resource constrained sensor nodes, furthermore, it successfully shifts the computation load and complexity to the BS. However, there is a trade-off between the detection error and computation complexity in the state-of-art CS-MUD algorithms. With the total number of sensor nodes growing and consequently the increase of the measurement matrix, the computation complexity of multiuser detection will increase dramatically which calls for a more efficient CS-MUD algorithm. The performance of CS based multiple access schemes is determined by the correlation of the non-orthogonal spreading sequences and the number of available sequences. Therefore, it is desirable to invent new spreading sequences with low cross-correlation. Combination of CS-MUD with other NOMA techniques will create new opportunities to utilize the advantages of different schemes and further improve the overloading and radio resource sharing for mMTC in the 5G communication system.\par



\bibliographystyle{unsrt}
\bibliography{bibfile}

\end{document}